\documentclass[fleqn,useAMS,usenatbib]{mnras}

\usepackage{amsmath}
\usepackage{amssymb}
\usepackage{graphicx}
\usepackage{times}
\usepackage{hyperref}
\usepackage{ulem}

\usepackage[flushleft]{threeparttable}

\usepackage{pifont}

\newcommand{\tickYes}{\ding{51}}
\newcommand{\tickNo}{\ding{55}}

\defcitealias{Liao2023}{L23}

\title[Impact of AGN feedback on coalescing SMBHs]
{RABBITS -- II. The impact of AGN feedback on coalescing supermassive black holes in disc and elliptical galaxy mergers}

\author[S. Liao et al.]
{Shihong Liao,$^{1,2}$\thanks{Email: shliao@nao.cas.cn} Dimitrios Irodotou,$^{1}$  Peter H. Johansson,$^{1}$ Thorsten Naab,$^{3}$ \newauthor Francesco Paolo Rizzuto,$^{1}$ Jessica M. Hislop,$^{1}$ Ruby J. Wright,$^{1}$ Alexander Rawlings$^{1}$
\\
$^1$Department of Physics, University of Helsinki, Gustaf Hällströmin katu 2, FI-00014 Helsinki, Finland\\
$^2$Key Laboratory for Computational Astrophysics, National Astronomical Observatories, Chinese Academy of Sciences, Beijing 100101, China\\
$^3$Max-Planck-Institut f\"ur Astrophysik, Karl-Schwarzchild-Str 1, D-85748 Garching, Germany\\
}

\begin{document}



\maketitle

\label{firstpage}

\begin{abstract}
In this study of the `Resolving supermAssive Black hole Binaries In galacTic hydrodynamical Simulations' (RABBITS) series, we investigate the orbital evolution of supermassive black holes (SMBHs) during galaxy mergers. We simulate both disc and elliptical galaxy mergers using the \textsc{ketju} code, which can simultaneously follow galaxy (hydro-)dynamics and small-scale SMBH dynamics with post-Newtonian corrections. With our SMBH binary subgrid model, we show how active galactic nuclei (AGNs) feedback affects galaxy properties and SMBH coalescence. We find that simulations without AGN feedback exhibit excessive star formation, resulting in merger remnants that deviate from observed properties. Kinetic AGN feedback proves more effective than thermal AGN feedback in expelling gas from the centre and quenching star formation. The different central galaxy properties, which are a result of distinct AGN feedback models, lead to varying rates of SMBH orbital decay. In the dynamical friction phase, galaxies with higher star formation and higher SMBH masses possess denser centres, become more resistant to tidal stripping, experience greater dynamical friction, and consequently form SMBH binaries earlier. As AGN feedback reduces gas densities in the centres, dynamical friction by stars dominates over gas. In the SMBH hardening phase, compared to elliptical mergers, disc mergers exhibit higher central densities of newly formed stars, resulting in accelerated SMBH hardening and shorter merger time-scales (i.e. $\la 500$ Myr versus $\ga 1$ Gyr). Our findings highlight the importance of AGN feedback and its numerical implementation in understanding the SMBH coalescing process, a key focus for low-frequency gravitational wave observatories.
\end{abstract}


\section{Introduction}\label{sec:intro}

Feedback from accreting supermassive black holes (SMBHs, with masses $M_{\rm BH}$ ranging from ${\sim} 10^{5}$ to ${\sim} 10^{10}$ ${\rm M}_{\sun}$), usually referred to as active galactic nuclei (AGNs) feedback, has become a vital ingredient in numerical simulations of galaxy formation and evolution \citep[see e.g.][for recent reviews]{Somerville2015,Naab2017,Crain2023}. The AGN feedback process is essential in reproducing the observed \citep[e.g.][]{Ferrarese2000,Gebhardt2000,Merloni2003,Haring2004,Kormendy2013,Saglia2016,Sahu2020,Baker2023} tight correlations between the properties of massive galaxies and their central SMBHs \citep[e.g.][]{DiMatteo2005,DiMatteo2008,Robertson2006,Habouzit2021}, quenching elliptical galaxies \citep[e.g.][]{DiMatteo2005,Springel2005feedback,Eisenreich2017}, reproducing the bright end of the observed galaxy luminosity function \citep[e.g.][]{Bower2006,Cattaneo2006,Croton2006,Somerville2008}, and it can also affect the structural and dynamical properties of galaxies \citep[e.g.][]{Bonoli2016,Dubois2016,Frigo2019,Valentini2020,Irodotou2022,Lokas2022,Mercedes-Feliz2023}.

Galaxy mergers naturally occur within the framework of hierarchical structure assembly in the cold dark matter scenario \citep*[see e.g.][for reviews]{Mo2010,Frenk2012}. When two massive galaxies merge, it is predicted that their central SMBHs undergo a three-phase evolutionary process \citep*{Begelman1980}: (i) at ${\sim}$kpc scales, the two SMBHs sink to the galaxy remnant centre due to dynamical friction from stars and gas \citep{Chandrasekhar1943,Ostriker1999}, eventually forming a gravitationally bound binary. (ii) At ${\sim}$pc scales, the SMBH binary continues to harden (i.e. its orbit shrinks) by losing energy and angular momentum to stars through gravitational slingshot interactions \citep*[e.g.][]{Mikkola1992,Quinlan1996,Sesana2006}. The SMBH binary will also interact with the circumbinary gas disc \citep[see][for a review]{Lai2022}. (iii) At ${\sim}$mpc scales, the binary evolution is dominated by gravitational wave (GW) emission \citep{Peters1963,Peters1964} until coalescence. The GW emitted from these coalescing SMBH binaries are key targets for the ongoing and upcoming low-frequency GW observatories, such as pulsar timing array \citep{Burke-Spolaor2019,Agazie2023} operating in the nano-hertz frequency range, and the future space GW observatories working in the milli-hertz frequency range, including the Laser Interferometer Space Antenna \citep[LISA,][]{Amaro-Seoane2017,Amaro-Seoane2023}, TianQin \citep{Luo2016}, and Taiji \citep{Ruan2020}.

AGN feedback plays an important role in shaping the general properties of merging galaxies. Early simulations using the thermal AGN feedback model \citep{Springel2005feedback} revealed that in gas-rich disc galaxy mergers, the intense nuclear gas inflows induced by tidal interactions trigger starbursts and provide fuel to the central SMBHs. Subsequently, the high SMBH accretion rates result in powerful AGN feedback which in turn expels the gas, reduces nuclear star formation, and quenches the galaxy remnant \citep[e.g.][]{DiMatteo2005,Springel2005feedback,Hopkins2006,Johansson2009}. Later simulations demonstrated that different implementations of AGN feedback can significantly impact specific galaxy properties \citep[e.g.][]{Choi2012,Choi2014,Choi2015,Debuhr2012,Wurster2013,Barai2014,Eisenreich2017}. For instance, compared to thermal AGN feedback, the kinetic (or mechanical) AGN feedback leads to higher-velocity gas outflows, reduced thermal gas X-ray luminosity, and greater AGN variability \citep[e.g.][]{Choi2012,Choi2014,Debuhr2012,Barai2014}. Recently, an AGN-driven wind injection model has been introduced to further improve the implementation of AGN feedback in galaxy formation simulations \citep[see e.g.][]{Costa2020,Torrey2020,Bollati2023b}.

However, one limitation of the aforementioned hydrodynamical galaxy merger simulations is the adoption of a simplified ad hoc merger criterion. In these simulations, two SMBHs are considered to merge instantaneously when their distance falls below a certain predefined threshold (e.g. the SMBH smoothing length and/or gravitational softening length), typically at the kpc or sub-kpc scales.\footnote{Here, the SMBH smoothing length is defined using the smoothed particle hydrodynamics approach. Note that in addition to the distance criterion, some simulations also incorporate a velocity criterion, i.e. the relative velocity between the two SMBHs should be less than either the local sound speed or the circular velocity.} This approach means that the three-phase coalescence process of SMBHs is not fully resolved in these simulations. Consequently, the GW-related predictions (e.g. SMBH merger rates) from cosmological galaxy formation simulations, which often employ similar SMBH merger criteria as galaxy merger simulations, rely on either an ad hoc constant delay time \citep[e.g.][]{Salcido2016,DeGraf2021} or post-processing semi-analytical models \citep[e.g.][]{Kelley2017,Volonteri2020,Chen2022,Li2022}.

Efforts have been made to improve the modelling of SMBH pair (i.e. two close SMBHs that have not yet formed a gravitationally bound binary) or binary dynamics in galactic-scale simulations. These improvements can be categorized into two groups: elliptical and disc galaxy mergers.

{\bf Elliptical galaxy mergers.} For gas-poor elliptical galaxies, the approximation of gas free is typically adopted and pure $N$-body simulations are used to study the SMBH coalescence process in such systems \citep[e.g.][]{Quinlan1997,Milosavljevic2001,Berczik2006,Khan2011,Rantala2019,Frigo2021,Nasim2021core}. Usually, the dynamical friction and binary hardening phases are resolved in these simulations and some of them also include post-Newtonian (PN) corrections to model the GW emission phase \citep[e.g.][]{Berentzen2009,Khan2012-EE-merger,Rantala2017,Mannerkoski2019}. From these collisionless simulations, it was found that the SMBH mergers typically take place on a ${\sim}$Gyr time-scale unless the SMBH binaries are highly eccentric \citep[e.g.][]{Khan2012-EE-merger,Vasiliev2015,Rantala2017}. 

However, all of these elliptical merger simulations do not consider gas and the associated physical processes (e.g. radiative cooling, SMBH accretion, and AGN feedback). In fact, even when we consider previous elliptical merger simulations that did not resolve small-scale SMBH dynamics \citep[e.g.][]{Gonzalez-Garcia2003,Boylan-Kolchin2005,Naab2006}, there is still a scarcity of literature on elliptical mergers that include gas \citep[for a few examples, see][]{Johansson2009,Sinha2009}. How does the presence of hot gas haloes in elliptical galaxies and the related physical processes impact the evolution of merging galaxies and coalescing SMBHs? Addressing this question can shed light on the validity of the gas-free approximation commonly adopted in previous elliptical merger simulations.

{\bf Disc galaxy mergers.} For gas-rich disc galaxies, the associated physical processes become significantly more complicated. Previous galactic-scale simulation studies which focused on the orbital decay of SMBHs can be grouped into two main categories \citep[see e.g.][for reviews]{Colpi2011,Dotti2012,Mayer2013}: (i) simulations of mergers involving two gas-rich disc galaxies \citep[e.g.][]{Kazantzidis2005,Mayer2007,Callegari2009,Callegari2011,Chapon2013,VanWassenhove2014,Roskar2015,Pfister2017,Chen2023} and (ii) simulations focusing on the evolution of two SMBHs within a circumnuclear gas disc which is regarded as the relic of a gas-rich galaxy merger \citep[e.g.][]{Escala2004,Escala2005,Dotti2006,Dotti2007,Dotti2009,Fiacconi2013,delValle2015,SouzaLima2017,SouzaLima2020,Bollati2023}. These simulations revealed that gas dynamical friction \citep{Escala2004,Escala2005,Dotti2006,Mayer2007,Chapon2013}, gravitational torques exerted on the binary from surrounding gas \citep{Escala2004,Escala2005,Dotti2006}, gas cooling and star formation which lead to more compact galaxy cores \citep{Callegari2009,Callegari2011,VanWassenhove2014,Chen2023}, and interactions with gas clumps \citep{Fiacconi2013,delValle2015,Roskar2015,SouzaLima2017} can assist the orbital decay of SMBHs in gas-rich environments.

However, the range of included physical processes varies across these disc galaxy simulations. Especially, only a handful of simulations \citep[e.g.][]{Callegari2011,VanWassenhove2014,Pfister2017,SouzaLima2017,Bollati2023,Chen2023} incorporate AGN feedback process. Recently, it was suggested that AGN feedback can weaken the gas dynamical friction on SMBHs \citep{Sijacki2011,SouzaLima2017,Bollati2023}. Moreover, AGN feedback can affect star formation and gas outflows in galaxies, thus altering the galactic structure and potential, which in turn should impact the motion of SMBH binaries. How exactly does AGN feedback, along with its numerical implementations, influence the SMBH orbital evolution in different phases? This question remains relatively unexplored and necessitates further investigation through additional simulation studies.

In addition, the above mentioned disc galaxy hydrodynamical simulations do not resolve the SMBH binary hardening and GW emission phases, i.e. the SMBH binaries or pairs in these simulations cease to shrink as they approach the adopted softening lengths (usually $\ga$ pc). Some attempts have been made to address the binary hardening and coalescence processes within gas-rich galaxy mergers. For example, \citet{Khan2016} selected a disc-disc galaxy merger from the Argo cosmological hydrodynamical simulation and resimulated the very central region of the galaxy remnant and the SMBH binary evolution using the direct $N$-body code {\sc $\phi$-gpu} with PN corrections. However, it is worth noting that their parent simulations did not include AGN feedback. Additionally, prior to the direct $N$-body calculations, the authors manually converted gas particles into stars \citep[see similar treatments in][]{Khan2012,Khan2018}. \citet{Koehn2023} extracted galaxies from the ROMULUS25 cosmological simulations and then resimulated the dynamics of SMBH triplets with direct $N$-body codes. The question then arises: How to resolve the SMBH binary dynamics in hydrodynamical simulations with a more self-consistent approach, avoiding these oversimplified approximations? This represents yet another challenge to be explored.

To address the questions mentioned above, we introduce the series of studies, {\it `Resolving supermAssive Black hole Binaries In galacTic hydrodynamical Simulations'} (RABBITS). In the RABBITS studies, we employ the {\sc ketju} code \citep{Rantala2017} to accurately model the dynamics of SMBH binaries in hydrodynamical simulations of galaxy mergers, including both disc and elliptical galaxies.

Based on the {\sc gadget-3} code \citep[last described by][]{Springel2005} which includes smoothed particle hydrodynamics (SPH) and galaxy formation subgrid models \citep{Hu2014,Eisenreich2017}, the {\sc ketju} code further incorporates the high-accuracy algorithmically regularized integrator {\sc mstar} \citep{Rantala2020} and PN corrections \citep{Blanchet2014}. This unique hybrid approach enables us to simultaneously capture galactic-scale galaxy formation processes and resolve the small-scale dynamics of SMBHs. The {\sc ketju} code has been successfully used to simulate the complex evolution of SMBHs in galaxy group environments in cosmological zoom-in simulations \citep{Mannerkoski2021,Mannerkoski2022}. Recently, \citet{Liao2023} \citepalias[hereafter][]{Liao2023} introduced a new accretion model for SMBH binaries to the {\sc ketju} code, which helps to improve the modelling of SMBH binaries in gas-rich galaxy mergers. Notably, \citetalias{Liao2023} incorporated both thermal and kinetic AGN feedback, allowing us to explore the impact of different implementations of AGN feedback.

Our focus in this paper (Paper II of the RABBITS series) is on studying how AGN feedback and its numerical implementations affect both galaxy properties and the orbital evolution of SMBH binaries. In a companion paper, \citet{Liao2023_PaperII} (hereafter Paper I), we focus on the impact of nuclear star formation on the hardening of SMBH binaries. The structure of this paper is as follows. In Section~\ref{sec:sim}, we describe the numerical code, the initial conditions, and the simulation details. The impact of AGN feedback on the properties of merging galaxies is presented in Section~\ref{sec:gal_prop}. In Section~\ref{sec:bh_orbit_decay}, we discuss the SMBH orbital decay in both the dynamical friction-dominated phase and the SMBH binary hardening and coalescence phase. We summarize and conclude in Section~\ref{sec:con}. In Appendices~\ref{ap:res_study} and \ref{ap:isolate_ETG}, we study the impact of numerical resolution and perform simulation tests of isolated elliptical galaxies.

\section{Numerical simulations}\label{sec:sim}

\subsection{Numerical code} \label{subsec:code}

All the simulations in this work are performed with the \textsc{ketju} code \citep{Rantala2017}, which is based on the \textsc{gadget-3} code \citep[i.e. the updated version of \textsc{gadget-2} described in][]{Springel2005} and incorporates the high-accuracy algorithmically regularized \textsc{mstar} integrator \citep{Rantala2020} to resolve the SMBH dynamics.\footnote{Note that a public \textsc{ketju} version based on the \textsc{gadget-4} code \citep{Springel2021}, which does not contain the galaxy formation subgrid model used in this work, was recently introduced in \citet{Mannerkoski2023}.} The code also includes PN corrections up to the order of 3.5PN for SMBH binaries \citep{Thorne1985,Mora2004,Blanchet2014}, which allows capturing the binary evolution down to the separation regime where the hardening process is predominantly driven by GW emission. The mass loss of the SMBH merger remnant due to GW emission is modelled using the fitting formulae from \citet{Zlochower2015}.

The \textsc{ketju} code uses the \textsc{sphgal} implementation \citep{Hu2014} for solving gas hydrodynamics. The adopted SPH kernel is the Wendland $C^4$ kernel and we set the number of SPH neighbours to $N_{\rm ngb} = 100$. To model the metal-dependent radiative cooling, star formation, and stellar feedback processes, the code adopts the subgrid model which was originally developed by \citet{Scannapieco2005,Scannapieco2006} and later improved in \citet{Aumer2013} and \citet{Nunez2017}. The cooling tables are adopted from \citet{Wiersma2009}. Stars form from gas particles in a stochastic implementation and the star formation criteria are: gas density $\rho_{\rm gas} \geq 2.2 \times 10^{-24}~{\rm g}~{\rm cm}^{-3}$ (or equivalently hydrogen number density $n_{\rm H} \geq 1~{\rm cm}^{-3}$), gas temperature $T_{\rm gas} \leq 1.2 \times 10^{4}~{\rm K}$, and the gas particle must be part of an convergent flow (i.e. the velocity divergence $\nabla \cdot \mathbfit{v}_{\rm gas} \leq 0$). The stellar feedback model includes the feedback from the Type Ia (SNIa) and Type II (SNII) supernovae explosions and the slow winds of asymptotic giant branch (AGB) stars.

For the SMBH subgrid model, we use the SMBH binary accretion model introduced in \citetalias{Liao2023}, which extends the widely used Bondi--Hoyle--Lyttleton (BHL) accretion \citep{Hoyle1939,Bondi1944,Bondi1952} model into the SMBH binary phase and incorporates the preferential accretion subgrid model of circumbinary discs \citep{Duffell2020}. Specifically, when an SMBH is in the single SMBH phase, its accretion rate is computed following the traditional BHL model \citep{Springel2005feedback}. \footnote{While the BHL model has been widely adopted in many galaxy formation simulations \citep[e.g.][]{Vogelsberger2013,Schaye2015,Pillepich2018}, there exists a considerable body of literature devoted to comparing the BHL estimation with accretion rates measured from high-resolution hydrodynamical simulations or proposing alternative accretion subgrid models \citep[e.g.][]{Hopkins2011,Power2011,Curtis2015,Curtis2016,Tremmel2017,Angles-Alcazar2021,Akerman2023}.} When two SMBHs form a bound binary, their total accretion rate is computed using the BHL formula and the gas properties at the centre of mass (CoM) of the SMBH binary, i.e.
\begin{equation}
    \dot{M}_{\rm bin} \equiv \dot{M}_{\rm BH,1} + \dot{M}_{\rm BH,2} =\alpha \frac{4 \upi G^2 M_{\rm bin}^2 \rho_{\rm CoM}}{(c_{\rm s, CoM}^2 + v_{\rm rel, CoM}^2)^{3/2}},
\end{equation}
where $\alpha = 25$ is the boost factor, $G$ is the gravitational constant, $M_{\rm bin} = M_{\rm BH,1} + M_{\rm BH,2}$ (assuming $M_{\rm BH,1} \geq M_{\rm BH,2}$) is the total mass of the SMBH binary, $M_{\rm BH,1}$ and $M_{\rm BH,2}$ are the masses of the primary SMBH and the secondary SMBH, $\rho_{\rm CoM}$ and $c_{\rm s, CoM}$ are the gas density and the sound speed at the CoM position of the binary, and $v_{\rm rel, CoM}$ is the magnitude of the relative velocity between the binary's CoM velocity and the gas velocity at the CoM position of the binary. The total accretion rate is capped by the Eddington limit of the binary system. Then, to distribute the total accretion rate between each SMBH of the binary system, we use the fitting formula from circumbinary disc simulations derived in \citet{Duffell2020}, \footnote{The circumbinary disc simulations in \citet{Duffell2020} do not incorporate AGN feedback. Therefore, adopting the fitting formula of Equation (\ref{eq:duffel2020}) implicitly assumes that AGN feedback does not influence the disc structure. However, AGN feedback could potentially impact the evolution of the SMBH binary + disc system \citep{delvalle2018}. As discussed in \citetalias{Liao2023}, this aspect can be improved in future work once fitting formulae from circumbinary disc simulations including AGN feedback are available.}
\begin{equation}\label{eq:duffel2020}
    \frac{\dot{M}_{\rm BH,2}}{\dot{M}_{\rm BH,1}} = \frac{1}{0.1 + 0.9q},
\end{equation}
where $q \equiv M_{\rm BH,2}/ M_{\rm BH,1} \in (0, 1]$ is the SMBH mass ratio. As a result, the secondary SMBH gets a higher accretion rate compared to the primary SMBH, which is called preferential accretion and drives the SMBH binary toward equal mass. This binary accretion model provides more physically motivated SMBH mass evolution in the binary phase, which is important for modelling the AGN feedback and the GW induced recoil velocities.

To study the impact of different AGN feedback implementations, we consider either pure thermal feedback or pure kinetic feedback in this work. For the thermal AGN feedback model, we follow the implementation in \citet{Springel2005feedback}. At each time-step, the total amount of energy given by
\begin{equation}
    \Delta E = \epsilon_{\rm f,thm} \epsilon_{\rm r} \dot{M}_{\rm BH} c^2 \Delta t
\end{equation}
is added to the internal energy of the surrounding gas particles according to their SPH kernel weights. Here, $\epsilon_{\rm r} = 0.1$ is the radiative efficiency, $\epsilon_{\rm f,thm} = 0.02$ is the thermal AGN feedbak efficiency, $c$ is the speed of light, and $\Delta t$ is the time-step.

For the kinetic AGN feedback model, we adopt an implementation similar to the kinetic mode in \citet{Weinberger2017}. At each time-step, the amount of feedback energy computed by
\begin{equation}
    \Delta E = \epsilon_{\rm f,kin} \epsilon_{\rm r} \dot{M}_{\rm BH} c^2 \Delta t
\end{equation}
is added to the SMBH feedback energy reservoir, $E_{\rm kin, res}$. Here $\epsilon_{\rm f,kin} = 0.008$ is the kinetic AGN feedback efficiency. Once the SMBH feedback energy reservoir reaches a given threshold, i.e.
\begin{equation}
    E_{\rm kin, res} \geq E_{\rm thr} = \frac{1}{2} f_{\rm thr} M_{\rm gas} \sigma_{\rm DM}^2,
\end{equation}
the energy in the reservoir is released to the surrounding $N_{\rm ngb}$ gas particles according to their SPH kernel weights. Here, $f_{\rm thr} = 20$ is the user-specified threshold parameter controlling the feedback strength, $M_{\rm gas}$ is the total gas mass within the SMBH smoothing length, and $\sigma_{\rm DM}$ is the one-dimensional velocity dispersion estimated from 40 nearby dark matter particles. The feedback energy, $E_{{\rm kin}, i}$, is coupled to the $i$-th gas neighbour in a kinetic form, i.e. this gas particle receives a kick velocity with the magnitude of
\begin{equation}
    v_{{\rm kick}, i} = \sqrt{\frac{2 E_{{\rm kin}, i}}{m_{{\rm gas},i}}},
\end{equation}
where $m_{{\rm gas}, i}$ is the mass of the $i$-th gas neighbour. In the non-binary phase, the direction of the kick velocity has an equal probability (50 per cent chance) of being either parallel or anti-parallel to  the direction of the gas particle's angular momentum, $\mathbfit{r}_{i} \times \mathbfit{v}_{i}$, where $\mathbfit{r}_{i}$ and $\mathbfit{v}_{i}$ are the position and velocity vectors of the gas particle with respect to the SMBH. In the binary phase, the kick direction has an equal 50 per cent probability of being parallel or anti-parallel to the direction of the binary orbital angular momentum.

Note that the thermal AGN feedback is implemented in a continuous approach, while the kinetic AGN feedback is in a pulsed manner. For more details of the galaxy formation subgrid model used in this study and the code implementations, we refer the interested reader to \citetalias{Liao2023}.

\subsection{Initial conditions}

\begin{figure*} 
\centering\includegraphics[width=425pt]{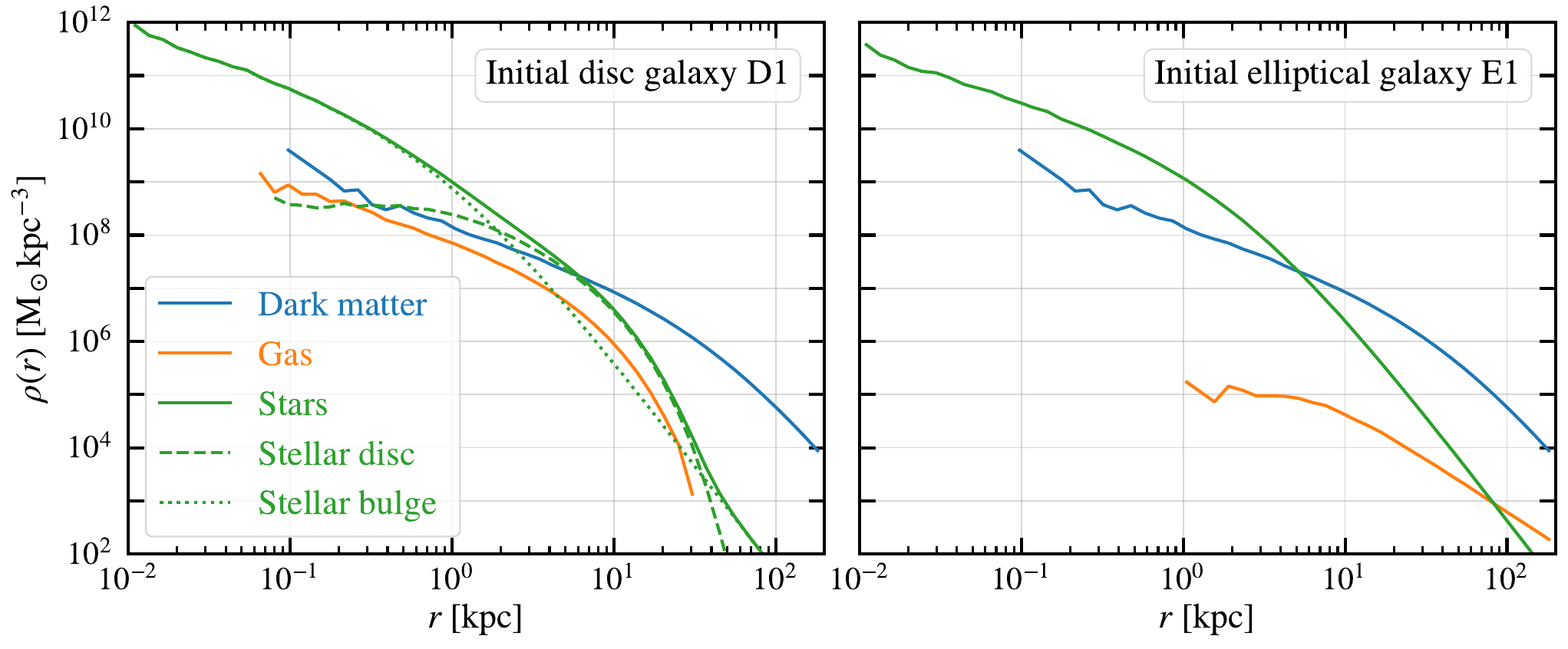}
\caption{Spherically averaged density profiles of the initial disc (D1, left) and elliptical (E1, right) galaxies. The dark matter, gaseous, and stellar components are plotted with blue, orange, and green, respectively. For D1, the density profiles of the stellar disc and stellar bulge are shown with the dashed and dotted curves, respectively. The two initial galaxies, D1 and E1, are designed to be `fraternal twins', i.e. they have identical virial masses, have similar masses in the dark matter, stellar, and gas components, but they differ in their morphology.}
\label{fig:init_rho}
\end{figure*}

\begin{table*}
\begin{threeparttable}
\caption{Properties of single progenitor galaxies. Both D1 and E1 galaxies are generated with a virial velocity of $V_{200} = 200~{\rm km}~{\rm s}^{-1}$, which is equivalent to a virial mass of $M_{\rm 200} = 2.62 \times 10^{12}~{\rm M}_{\sun}$ and a virial radius of $R_{200} = 282~{\rm kpc}$. Both galaxies contain an SMBH in the centre with a mass of $M_{\rm BH} = 7.5 \times 10^{7}~{\rm M}_{\sun}$. From the left, the name of the progenitor galaxy, the dark matter mass $M_{\rm DM}$, the stellar disc mass $M_{\star, {\rm disc}}$, the stellar bulge mass $M_{\star, {\rm bulge}}$, the gas disc mass $M_{\rm gas, disc}$, the gas halo mass $M_{\rm gas, halo}$, the dark matter particle number $N_{\rm DM}$, the stellar disc particle number $N_{\star, {\rm disc}}$, the stellar bulge particle number $N_{\star, {\rm bulge}}$, the gas disc particle number $N_{\rm gas, disc}$, and the gas halo particle number $N_{\rm gas, halo}$.}
\label{tab:ic_params}
\begin{tabular}{lccccccccccccc}
\hline
Galaxy & $M_{\rm DM}$ & $M_{\star, {\rm disc}}$ & $M_{\star, {\rm bulge}}$ & $M_{\rm gas, {\rm disc}}$ & $M_{\rm gas, {\rm halo}}$ & $N_{\rm DM}$ & $N_{\star, {\rm disc}}$ & $N_{\star, {\rm bulge}}$ & $N_{\rm gas, {\rm disc}}$ & $N_{\rm gas, {\rm halo}}$\\
 & $[10^{10}~{\rm M}_{\sun}]$ & $[10^{10}~{\rm M}_{\sun}]$ & $[10^{10}~{\rm M}_{\sun}]$ & $[10^{10}~{\rm M}_{\sun}]$ & $[10^{10}~{\rm M}_{\sun}]$ & $[10^5]$ & $[10^5]$ & $[10^5]$ & $[10^5]$ & $[10^5]$\\
\hline
D1 & 247.66 & 8.59 & 3.58 & 2.15 & -- & 16.00 & 8.60 & 3.58 & 2.15 & -- \\
E1 & 247.17 & -- & 12.17 & -- & 2.64 & 16.00 & -- & 12.18 & -- & 2.64\\
\hline
\end{tabular}
\end{threeparttable}
\end{table*}

\begin{table}
\begin{threeparttable}
\caption{Initial conditions of galaxy mergers.}
\label{tab:sim_params}
\begin{tabular}{lccccccccccccc}
\hline
Parameters & DM & Stars & Gas & SMBH\\
\hline
Particle mass $[10^{5}{\rm M}_{\sun}]$ & $15.5$ & $1$ & $1$ & $750$\\
Particle number & $3.2$M & $2.4$M & ${\sim} 0.5$M$^\dagger$ & $2$ \\
Softening length $[{\rm pc}]$ & $100$ & $5$ & $20$ & $5$\\
\hline
\end{tabular}
\begin{tablenotes}[flushleft]
 \footnotesize
 \item $\dagger$ The initial total gas particle number is $4.30 \times 10^5$ for the DD-11-G5 run, and $5.28 \times 10^5$ for the EE-11-G5 run. Note that `M' in this row denotes a factor of one million ($10^{6}$).
\end{tablenotes}
\end{threeparttable}
\end{table}

\begin{table*}
\begin{threeparttable}
\caption{Simulation sets including different physical processes. The symbols `\tickYes' (Yes) and `\tickNo' (No) are used to denote whether a physical process is included in a simulation. Here, the hydrodynamics is solved using the SPH method. The radiative cooling process takes metal-dependent cooling into account. Star formation criteria are based on the density and temperature thresholds. Stellar feedback includes the feedback from SNIa, SNII, and AGB stars. SMBH accretion is modelled using the BHL accretion and the preferential accretion from circumbinary discs. The thermal AGN feedback is performed by increasing the internal energy of the surrounding gas particles, while the kinetic AGN feedback is implemented by adding velocity kicks to the ambient gas. Further details regarding various physical processes are provided in Section~\ref{subsec:code}.}
\label{tab:sim_set}
\begin{tabular}{lccccccc}
\hline
Name & Gravity & Hydrodynamics & Radiative & Star formation & SMBH & Thermal & Kinetic \\
 & \& PN corrections & (SPH) & cooling & \& stellar feedback & accretion & AGN feedback & AGN feedback \\
\hline
NoGas & \tickYes & \tickNo & \tickNo & \tickNo & \tickNo & \tickNo & \tickNo \\
NoCool & \tickYes & \tickYes & \tickNo & \tickNo & \tickNo & \tickNo & \tickNo \\
CoolStarNoAGN & \tickYes & \tickYes & \tickYes & \tickYes & \tickNo & \tickNo & \tickNo \\
CoolStarThmAGN & \tickYes & \tickYes & \tickYes & \tickYes & \tickYes & \tickYes & \tickNo \\
CoolStarKinAGN & \tickYes & \tickYes & \tickYes & \tickYes & \tickYes & \tickNo & \tickYes \\
\hline
\end{tabular}
\end{threeparttable}
\end{table*}

In this study, we consider both disc-disc and elliptical-elliptical galaxy mergers. To generate the initial conditions, we first prepare the single progenitor galaxies. For the disc progenitor galaxy, we use the D1 galaxy from \citetalias{Liao2023} but with an initial SMBH mass of $M_{\rm BH} = 7.5 \times 10^7~{\rm M}_{\sun}$, which matches the observed SMBH mass--stellar velocity dispersion ($M_{\rm BH}$--$\sigma_\star$) relation from \citet{Kormendy2013}, instead of a seed SMBH mass ($10^5~{\rm M}_{\sun}$)\footnote{The motivation of starting with a seed SMBH mass in \citetalias{Liao2023} was to test the SMBH accretion and feedback models for cosmological simulations.}. The SMBH spin is set to zero, and we do not model the evolution of the SMBH spin due to gas accretion in this study. The D1 disc progenitor is created following the method in \citet{Springel2005feedback}. It consists of a \citet{Hernquist1990} dark matter halo (with a concentration of $c_{200} = 9$ and a spin parameter of $\lambda = 0.033$), an exponential disc (with a scale length $r_{\rm d}^{\rm D1} = 4.32$ kpc, a mass fraction $m_{\rm d}^{\rm D1} = 0.041$, and a gas fraction $f_{\rm gas}^{\rm D1} = 0.2$), a Hernquist bulge (with a scale length $r_{\rm b}^{\rm D1} = 0.2 r_{\rm d}^{\rm D1}$ and a mass fraction $m_{\rm b}^{\rm D1} = m_{\rm d}^{\rm D1}/3$), and a central SMBH. The initial stellar ages and stellar/gaseous metallicities are set following \citet{Lahen2018}. The density profiles of the different components of D1 are shown in the left panel of Fig.~\ref{fig:init_rho}; the component masses and particle numbers are summarized in Table~\ref{tab:ic_params}. We refer the reader to \citetalias{Liao2023} for more details.

The elliptical progenitor galaxy, which is denoted as E1 here, is designed to be a `fraternal twin' of D1, i.e. they have similar masses/sizes in different components but differ in galaxy morphology. Such controlled numerical setups enable us to study the impact of galaxy morphology on SMBH mergers. Following the setup of D1, we use the same virial velocity $V_{200} = 200~{\rm km}~{\rm s}^{-1}$ to generate E1, which leads to the same virial mass ($M_{200} = 2.62 \times 10^{12}~{\rm M}_{\sun}$) and virial radius\footnote{Here, the virial radius is defined as the radius within which the mean matter density is 200 times the cosmic critical density at $z=0$.} ($R_{200} = 282~{\rm kpc}$). The dark matter halo of E1 is similar to that of D1. For the stellar component, E1 only has a Hernquist bulge, and its total stellar mass is set to be the sum of the stellar disc and bulge masses of D1, and its projected effective radius is set to $r_{\rm b, proj}^{\rm E1} = r_{\rm d}^{\rm D1} = 4.32~{\rm kpc}$, which corresponds to a 3D scale radius of $r_{\rm b}^{\rm E1} = 2.38~{\rm kpc}$. The ages and metal abundances of star particles are initialized in the same manner as for the D1 bulge component. For the gaseous component, E1 has a hot gas halo with the mass within $R_{200}$ equal to the gas disc mass of D1, i.e. $M_{200}^{\rm gas} = 2.15 \times 10^{10}~{\rm M}_{\sun}$. The hot gas halo follows the $\beta$-model profile, which provides a good fit to the hot gas around elliptical galaxies \citep{Forman1985}. We adopt an outer slope parameter $\beta = 2/3$ \citep{Jones1984} and a core radius of $r_{\rm c} = 0.22 r_{\rm s} = 6.9~{\rm kpc}$ \citep{Makino1998}, where $r_{\rm s} \equiv R_{200} / c_{200} = 31.3~{\rm kpc}$ is the scale radius of the dark matter halo. The cutoff radius of the gas halo, beyond which there are no gas particles, is set to $r_{\rm cut} = 50 r_{\rm c} = 345~{\rm kpc}$. The total mass of the gas halo is $M_{\rm gas, halo} = 2.64 \times 10^{10}~{\rm M}_{\sun}$. We assume no angular momentum transfer between the dark matter halo and the gas halo, and thus the gas halo has the same specific angular momentum as the dark matter halo. The radial metallicity profile of the gas halo is assumed to follow the stellar one. Following D1, E1 also contains a central SMBH with a mass of $M_{\rm BH} = 7.5 \times 10^7~{\rm M}_{\sun}$ and a spin of zero. Note that \citet{Eisenreich2017} adopted a similar approach to set up their initial elliptical galaxy. The density profiles of E1 are shown in the right panel of Fig.~\ref{fig:init_rho}; the masses and particles numbers of different components are summarized in Table~\ref{tab:ic_params}.

We have compared the properties of D1 and E1 to the observational data, and confirmed that both galaxies agree reasonably well with the observations; see Section~\ref{sec:gal_prop} for details. The D1 and E1 progenitor galaxies are then used to create the initial conditions of equal-mass galaxy mergers. For both disc-disc and elliptical-elliptical galaxy mergers, two identical galaxies are set to approach each other on a parabolic orbit with an initial separation of $d_{\rm sep, ini} = R_{200}$ and a pericentric distance of $d_{\rm peri, ini} = 2r_{\rm d}^{\rm D1}$. We adopt the G5 retrograde galaxy orbit geometry from \citet{Naab2003} for the disc-disc galaxy merger, resulting in a relatively modest starburst during the merger and making the simulation computationally efficient. For the G5 orbit, the inclination of the disc with respect to the orbital plane, $i$, and the argument of pericentre, $\omega$, of the two galaxies are $i_{1} = -109^\circ$, $\omega_{1} = -60^\circ$, $i_{2} = 180^\circ$, $\omega_{2} = 0^\circ$. Note that here the definitions of $i$ and $\omega$ follow the convention in \citet{Toomre1972}. Following the philosophy of fraternal twin simulations, for the elliptical-elliptical galaxy merger, as the rotation of the progenitor galaxy (e.g. the dark matter halo and the gas halo) introduces a special direction, we set the initial galaxy orbit also as G5 by replacing the disc orientation with the galaxy spin direction and the disc plane with the halo equatorial plane.

Following the naming convention of galaxy mergers in \citetalias{Liao2023}, i.e. `progenitor galaxy types-galaxy mass ratio-orbit geometry', we denote our disc-disc and elliptical-elliptical galaxy mergers as DD-11-G5 and EE-11-G5, respectively. The particle masses and the total particle numbers of the simulation initial conditions are summarized in Table~\ref{tab:sim_params}.

\subsection{Simulations}\label{subsec:sim}

To study how different physical processes (especially AGN feedback) affect the properties of merging galaxies and the orbital decay of SMBH binaries, for both DD-11-G5 and EE-11-G5 galaxy mergers, we perform five simulation sets including different physical processes and model implementations:

(i) {\bf NoGas}. This simulation set is gravity-only. In the initial condition, all gas particles are converted into stars by inheriting the positions, velocities, and masses. This set of simulations is an analogue to the collisionless gas-free elliptical galaxy mergers in previous studies. Note that here we do not refer to these simulations as `collisionless' runs because they do resolve the SMBH--SMBH and the SMBH--star collisional interactions.

(ii) {\bf NoCool}. Compared to the NoGas set, this set of simulations includes not only gravity but also gas hydrodynamics with an adiabatic equation of state. These runs do not incorporate the radiative cooling process, and as a result, the processes of star formation, as well as stellar and AGN feedback, are not taken into account. Note that similar runs have been referred to as `adiabatic' or `non-radiative' in the literature.

(iii) {\bf CoolStarNoAGN}. Compared to the NoCool set, these simulations further include the subgrid models of gas cooling, star formation, and stellar feedback. However, SMBH accretion and AGN feedback processes are not considered in these simulations.

(iv) {\bf CoolStarThmAGN}: Compared to the CoolStarNoAGN set, this set of simulations includes in addition the SMBH binary accretion model and the thermal AGN feedback mode.

(v) {\bf CoolStarKinAGN}: These simulations are similar to the CoolStarThmAGN ones, but use the kinetic AGN feedback mode instead of the thermal mode.

A summary of the physical processes included in the different simulation sets is given in Table~\ref{tab:sim_set}. All simulations in this work start from the cosmic time of $t_0 = 10.7~{\rm Gyr}$, with most evolving for 3 Gyr. However, the SMBH binaries in the DD-11-G5 NoGas and the EE-11-G5 CoolStarThmAGN runs take longer to merge, so these two simulations are evolved for 4 Gyr. For a straightforward comparison with the EE-11-G5 CoolStarThmAGN run, the EE-11-G5 CoolStarKinAGN run is also evolved for 4 Gyr, despite the fact that the SMBH merger takes place at ${\sim} 2.95$ Gyr.

Following \citetalias{Liao2023}, the Plummer-equivalent gravitational softening lengths for dark matter, gas, star, and SMBH particles are set to $100$ pc, $20$ pc, $5$ pc, and $5$ pc, respectively. Note that only the SMBH--star and SMBH--SMBH interactions are unsoftened in our simulations, the SMBH--gas and SMBH--dark matter gravitational interactions are still softened. The radius of the KETJU region around each SMBH is $15$ pc, i.e. three times the stellar/SMBH softening length. Compared to the runs in \citetalias{Liao2023} which start with a seed SMBH mass and the KETJU integration is only switched on when the SMBH-to-star particle mass ratio reaches the threshold $M_{\rm BH}/m_{\star} = 300$, the simulations in this work have an initial mass ratio of $M_{\rm BH}/m_{\star} = 750$, and the KETJU integration is switched on from the very beginning. In this paper, the simulation time, $t$, is defined as the time span after $t_0$. The CoolStarThmAGN and CoolStarKinAGN simulations utilize the same numerical setup and parameters as in \citetalias{Liao2023}, thus we refer the interested reader to \citetalias{Liao2023} for further simulation details.

\section{Galaxy properties}\label{sec:gal_prop}

\subsection{Overview of galaxy evolution} \label{subsec:overview}
\begin{figure*} 
\centering\includegraphics[width=450pt]{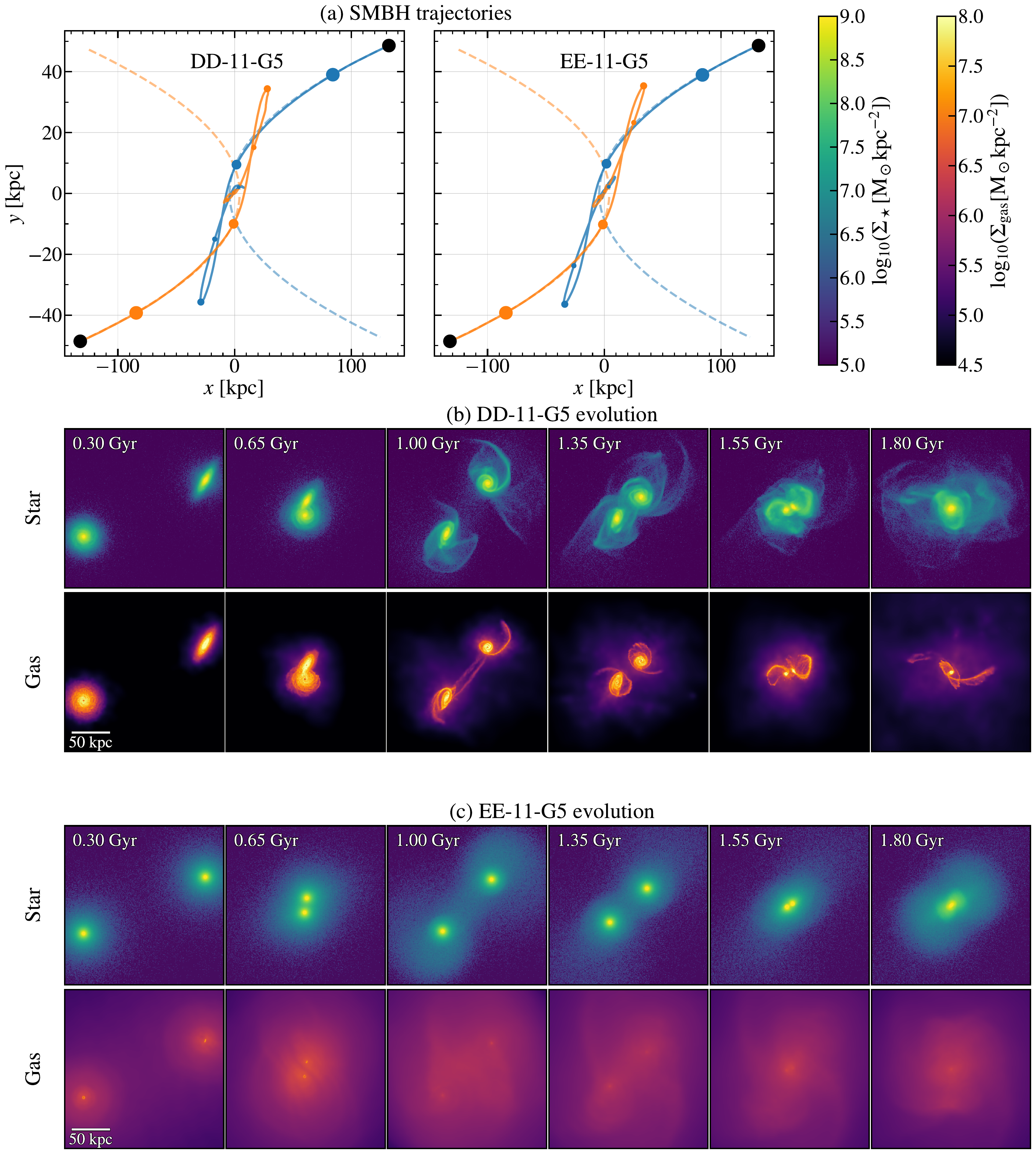}
\caption{SMBH trajectories and the evolution of the stellar and gaseous components in the CoolStarKinAGN simulations. (a) SMBH trajectories for both the DD-11-G5 (left) and EE-11-G5 (right) runs. The incoming parabolic orbits are plotted with dashed curves while the actual SMBH paths are shown with solid curves. The black circles mark the initial SMBH positions. The coloured circles mark the positions at six later snapshots, with smaller sizes representing later time. The six corresponding snapshots of the galaxy evolution are plotted in Panels b and c. (b) Evolution of the projected stellar (upper) and gaseous (lower) densities from the DD-11-G5 simulation. During the merger of disc galaxies, stellar/gaseous tails and bridges launch as a result of the tidal response. The colour bars for the projected densities are given at the top-right corner of the figure. (c) Similar to Panel b, but for the EE-11-G5 run. The same colour bars as in the DD-11-G5 case have been adopted here. Unlike the disc galaxy mergers, no narrow tidal tails or bridges are observed in the elliptical galaxy mergers.}
\label{fig:trajectory_visua}
\end{figure*}

We start by looking at the overall galaxy merger process. In Fig.~\ref{fig:trajectory_visua}, we use the CoolStarKinAGN runs of both the DD-11-G5 and EE-11-G5 mergers as examples to illustrate the SMBH trajectories and the galaxy interactions from our simulations. 

The left plot of Panel (a) and the plots in Panel (b) of Fig.~\ref{fig:trajectory_visua} depict the evolution of the DD-11-G5 galaxy merger. Initially the two galaxies are set on parabolic orbits (i.e. the dashed lines in Panel (a)) with a separation of $R_{200}$. Within the first ${\sim} 650$ Myr (i.e. prior to the first pericentre passage), while the extended dark matter haloes of the two galaxies are already interacting with each other, the central regions of the galaxies, including the central SMBHs, remain tightly bound and experience minimal tidal interactions. As a result, they continue to follow the input parabolic orbits and move ahead of their own dark matter haloes. When the galaxy centres get closer and go through the first pericentre passage, the intense tidal interactions between discs deform their appearances and produce elongated and narrow tails and bridges \citep{Toomre1972}. The gaseous tidal features tend to be narrower and better defined compared to the stellar ones, reflecting that radiative cooling in gas helps to damp the random motions. At the same time, the strong gravitational pulls from the lagging dark matter haloes slow down the galaxy centres. At $t \sim 1$ Gyr, the galaxy centres reach their first orbital apocentre and fall back. During this process, the tidal forces convert the orbital energy and angular momentum into the internal degrees of freedom in each galaxy, and consequently the galaxy centres deviate from the initial parabolic orbits. The centres of galaxies undergo a few passages, and with each cycle, the associated tidal interactions increasingly stimulate the internal motions of the inner regions. As a result of the continuous loss of orbital energy and angular momentum, the two galaxies/SMBHs eventually merge together and the galaxy merger remnant continues to undergo relaxation.

The overall orbital evolution of the EE-11-G5 case is qualitatively similar as the aforementioned disc system. At the early stage, the galaxy centres also follow the initial parabolic orbits before their first pericentre passage. However, unlike the cold disc system, the spherical elliptical galaxies do not exhibit narrow tidal tails or bridges, instead, we only observe broad features, i.e. some stars are excited to wider orbits by the tidal interactions and the overall stellar distribution is more extended. The physical reason behind these broad tidal features is that the random motions of individual stars in such dispersion-supported systems are comparable to the velocities imparted by the tidal forces. The initial spherical hot gas haloes also undergo deformation due to tidal torques and particularly the additional gas pressure, leading to the dispersal of their high-density cores.

\subsection{Star formation and SMBH growth histories}

\begin{figure*} 
\centering\includegraphics[width=400pt]{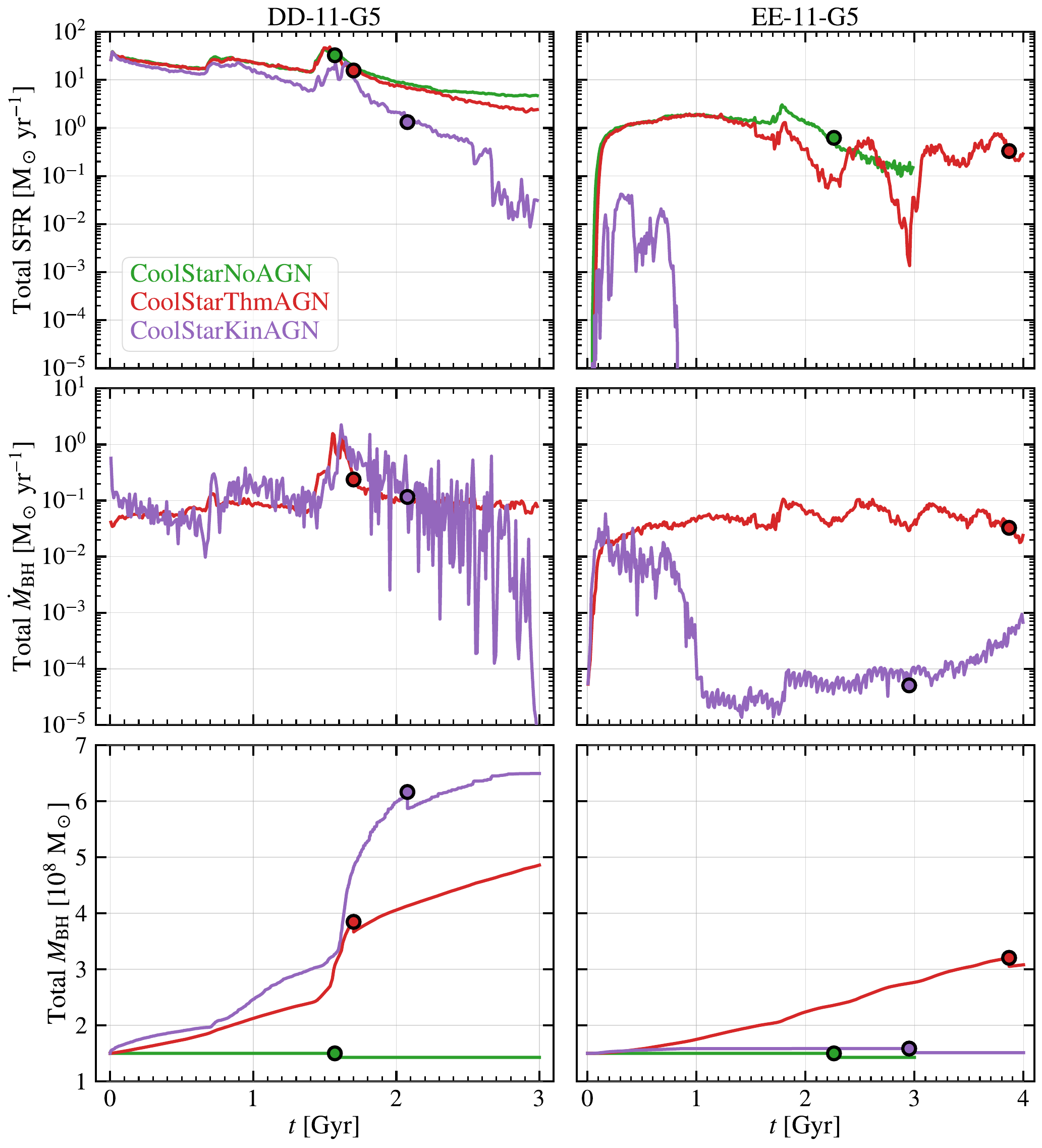}
\caption{Star formation and SMBH accretion histories in the DD-11-G5 (left) and EE-11-G5 (right) mergers. From top to bottom, the time evolution of the total SFRs, the total SMBH accretion rates, and the total SMBH masses in the simulations are plotted. The CoolStarNoAGN, CoolStarThmAGN, and CoolStarKinAGN runs are displayed with green, red, and purple lines, respectively. The filled circles mark the SMBH merger events in the different runs. Note that the SFRs and the SMBH accretion rates are averaged over 10 Myr here. The kinetic AGN feedback implementation is more effective in suppressing the star formation in galaxies and it results in more bursty SMBH accretion rates.}
\label{fig:sfr_bhar}
\end{figure*}

\begin{figure*} 
\centering\includegraphics[width=\textwidth]{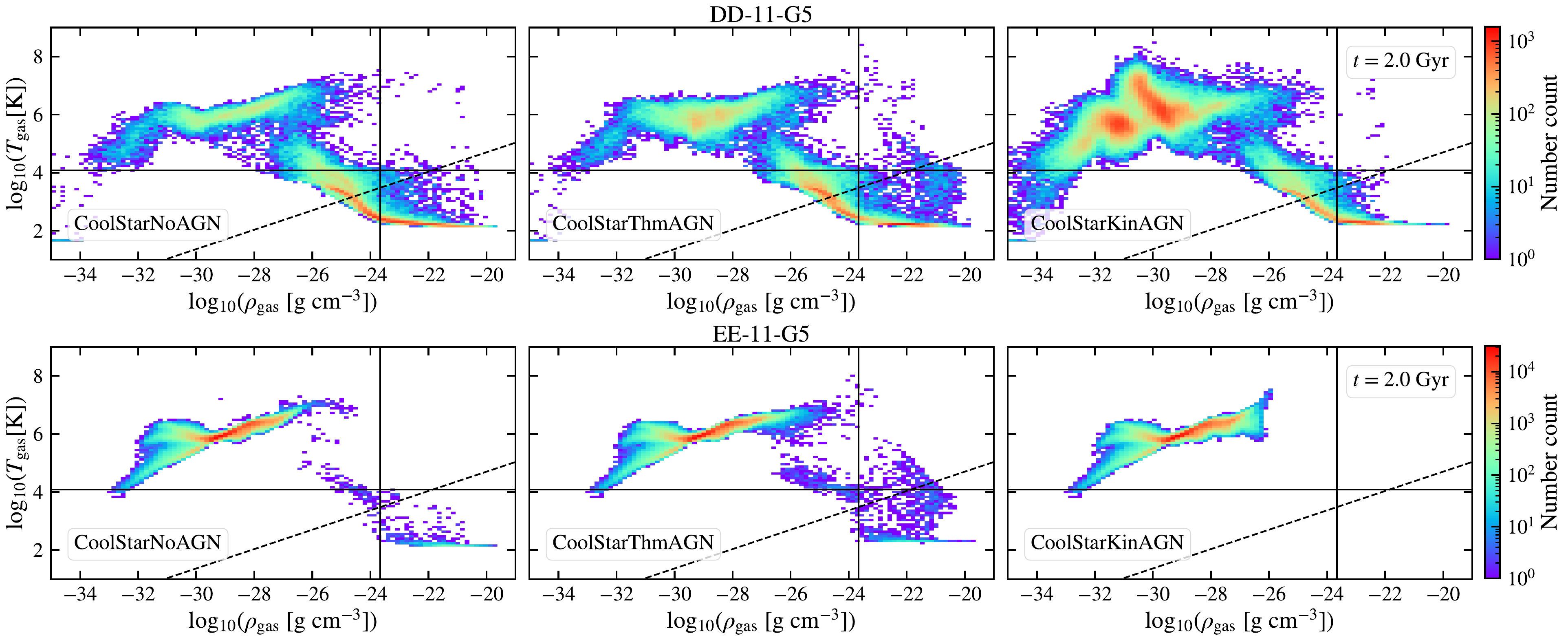}
\caption{Gas phase diagrams at $t=2$ Gyr. The DD-11-G5 and the EE-11-G5 runs are shown in the top and bottom panels, respectively. From left to right, the CoolStarNoAGN, CoolStarThmAGN, and CoolStarKinAGN runs are plotted. The colour represents the number of gas particles in the different temperature--density bins, and the colour bar for each row is displayed on the right. The horizontal and the vertical solid lines mark the temperature and the density thresholds for star formation, respectively. The dashed line marks the Jeans mass resolution limit, $M_{\rm J, lim} = N_{\rm ngb} m_{\rm gas}$. For gas particles below this dashed line, the Jeans mass is not resolved and they can collapse under self-gravity due to numerical noise \citep[see][]{Eisenreich2017}. Compared to the cases of no AGN feedback and thermal AGN feedback, the kinetic AGN feedback model results in distinct gas distributions in the phase diagram, i.e. less gas particles in the star-forming phase and more gas in the hot and diffuse phase.}
\label{fig:phase_diagram}
\end{figure*}

In Fig.~\ref{fig:sfr_bhar}, we present the star formation and SMBH growth histories for the CoolStarNoAGN, CoolStarThmAGN, and CoolStarKinAGN runs, for both disc and elliptical galaxy mergers. Note that the NoGas and NoCool runs do not include the processes of star formation and SMBH mass growth, hence they are excluded from this figure.

In the top panels, the total star formation rates (SFRs) in the simulations as a function of time are plotted. The CoolStarNoAGN runs consistently exhibit the highest SFRs throughout the entire simulation for both the DD-11-G5 and EE-11-G5 mergers. However, once thermal AGN feedback is included, star formation is suppressed, particularly after $t \sim 1.5$ Gyr, when the SMBHs exhibit high accretion rates and produce strong AGN feedback. Notably, the implementation of kinetic AGN feedback proves to be more effective in suppressing star formation compared to thermal AGN feedback. In the case of the DD-11-G5 merger, the CoolStarKinAGN run exhibits a lower total SFR compared to the CoolStarThmAGN run throughout the entire simulation period. In the last ${\sim}0.5$ Gyr, the SFR of the CoolStarKinAGN merger remnant becomes approximately two orders of magnitude lower than that of the CoolStarThmAGN remnant. In the EE-11-G5 merger, the CoolStarKinAGN run results in a complete quenching of the SFR in both merging galaxies after the first pericentre passage ($t \sim 0.7$ Gyr). In contrast, the CoolStarThmAGN run maintains an SFR $\ga 0.1~{\rm M}_{\sun}{\rm yr}^{-1}$ for most of the simulation duration. These results are in agreement with previous simulation studies which also found that kinetic AGN feedback is more efficient in removing gas from the galaxy centre and thus quenching galaxies (e.g. \citealt{Choi2012,Choi2014,Barai2014,Eisenreich2017}; see also section 5.1 of \citealt{Costa2020} for a detailed discussion).

The impact of AGN feedback on gas can be further illustrated in the gas phase diagram. In Fig.~\ref{fig:phase_diagram}, we show the gas temperature--density phase diagrams at $t = 2$ Gyr, when the SMBHs either have recently merged or have formed a bound binary. The horizontal and vertical solid lines mark the star formation temperature and density thresholds, respectively. The gas particles in the bottom-right corner are star-forming. Compared to the no AGN case, the thermal AGN feedback heats up more gas particles in the dense region (i.e. $\rho_{\rm gas} \ga 10^{-24}~{\rm g}~{\rm cm}^{-3}$). In contrast, the kinetic AGN feedback kicks these gas particles with high velocity, causing them to leave the dense region and collide with gas particles at larger radii, thereby heating the outer diffuse gas. Consequently, the CoolStarKinAGN run tends to have more gas particles in the hot and diffuse phase compared to the no AGN feedback and the thermal AGN feedback cases. In the EE-11-G5 remnant, the kinetic AGN feedback is so effective that no star-forming gas is present, leading to a fully quenched galaxy remnant.

The evolution of the total SMBH accretion rates and the total SMBH masses are displayed in the middle and bottom panels of Fig.~\ref{fig:sfr_bhar}, respectively. In the DD-11-G5 case, the thermal and kinetic AGN feedback runs show similar overall trends in SMBH accretion histories. However, the CoolStarKinAGN run, with its pulsed feedback, displays more bursty SMBH accretion (or equivalently higher AGN variability) compared to the continuous thermal feedback. Following the merger of the two SMBHs, we observe a slight decrease in the total SMBH mass, attributed to the energy loss through GW emission \citep{Zlochower2015}. In the EE-11-G5 case, after the first pericentre passage, the CoolStarKinAGN run exhibits SMBH accretion rates approximately two orders of magnitude lower than those of the CoolStarThmAGN run. Consequently, the total SMBH mass in the CoolStarKinAGN run shows minimal growth, reaching only half of that in the CoolStarThmAGN run by the end of the simulation ($t = 4$ Gyr). This low SMBH accretion rate in the kinetic AGN feedback run can be understood as follows: unlike the cold gas reservoir in disc galaxies, the gas from the extended hot gas halo in elliptical galaxies must first cool down and reach the galaxy centre to be accreted by the SMBH. Since kinetic AGN feedback effectively maintains the gas heated to a large distance, preventing the formation of cooling flows and keeping a low gas density in the galaxy centre, the central SMBH experiences a considerably lower accretion rate.

We have conducted a resolution study to address the reliability of our results by running simulations with both higher and lower mass resolutions. This study confirms that the star formation and SMBH growth histories presented above have reached a satisfactory resolution convergence. See Appendix~\ref{ap:res_study} for details.

From the comparison among the CoolStarNoAGN, CoolStarThmAGN, and CoolStarKinAGN runs, we can clearly see that the presence and specific numerical implementation of AGN feedback have significant impact on star formation and SMBH growth. The kinetic AGN feedback is necessary to efficiently quench galaxy merger remnants, which typically become elliptical galaxies. In Appendix~\ref{ap:isolate_ETG}, we further test the AGN feedback implementations in maintaining a quiescent elliptical galaxy in isolation, and the results again emphasize the importance of kinetic AGN feedback in generating a red and quiescent early-type galaxy.

\subsection{Galaxy scaling relations}

\begin{figure*} 
\centering\includegraphics[width=400pt]{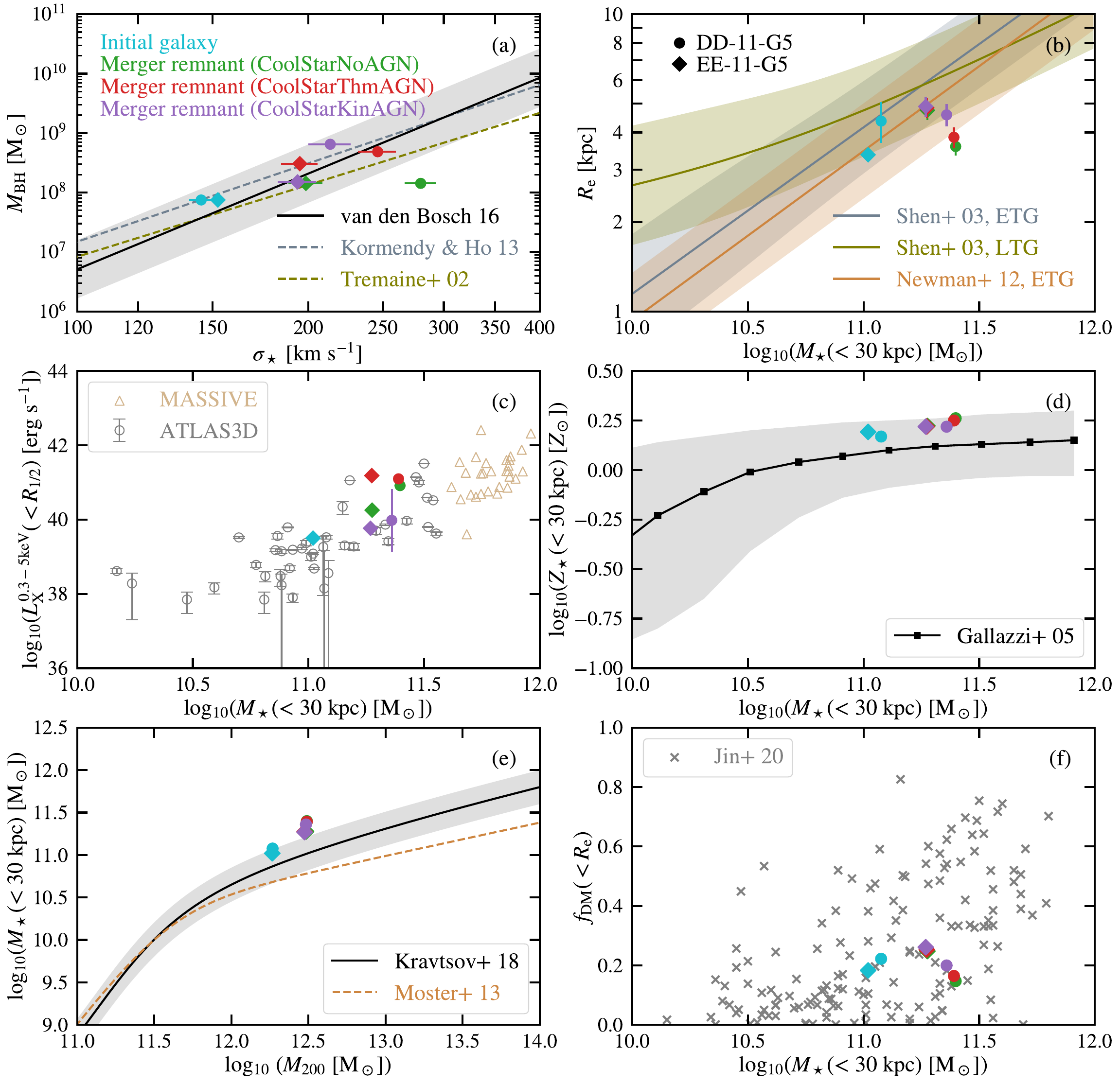}
\caption{Comparison of galaxy properties in simulations and observations. In all panels, the initial isolated galaxies are plotted with cyan, while the merger remnants in the CoolStarNoAGN, CoolStarThmAGN, and CoolStarKinAGN runs are shown with green, red, and purple, respectively. The DD-11-G5 and EE-11-G5 runs are plotted with filled circles and diamonds, respectively. (a) $M_{\rm BH}$--$\sigma_\star$ relation. The error bars in the line-of-sight stellar velocity dispersion show the standard deviations computed from 50 random projected orientations. The solid line plots the best-fitting $M_{\rm BH}$--$\sigma_\star$ relation from \citet{vandenBosch2016} and the grey-shaded regions show the intrinsic one-sigma scatter. The dashed lines give the best-fitting relations from \citet{Kormendy2013} and \citet{Tremaine2002}. (b) Galaxy size--stellar mass relation. Here the galaxy size is characterized using the projected half stellar mass radius $R_{\rm e}$. The solid lines and shaded regions show the observed relations for early-type galaxies (ETGs) and late-type galaxies (LTGs) from \citet{Shen2003} and \citet{Newman2012}. (c) The luminosity of soft X-rays (0.3--5 keV) from hot and diffuse gas versus the galaxy stellar mass. The X-ray luminosity is computed following the method detailed in \citetalias{Liao2023}. Note that the initial disc galaxy is not shown in this panel because it only has a cold gas disc and its hot gas X-ray luminosity is thus zero. The observational data from the MASSIVE \citep{Ma2014} and ATLAS3D \citep{Cappellari2011} surveys are from \citet{Goulding2016}. (d) Stellar metallicity--stellar mass relation. The black line and the grey-shaded region show the median observed relation and the 16th/84th percentiles from \citet{Gallazzi2005}. Here, we adopt $Z_{\sun} = 0.0127$ for the solar metallicity. (e) Stellar mass--halo mass relation. The black curve shows the best-fitting relation from \citet{Kravtsov2018} and the grey-shaded regions plot the scatter in stellar mass (i.e. 0.2 dex). The dashed line gives the best-fitting relation from \citet{Moster2013}. (f) Dark matter mass fraction within the projected effective radius as a function of galaxy stellar mass. The crosses plot the dark matter fractions of 149 MaNGA ETGs from \citet{Jin2020}. Overall, the galaxy merger remnants in the simulations with AGN feedback included agree well with the observations.}
\label{fig:gal_prop}
\end{figure*}

In Fig.~\ref{fig:gal_prop}, we provide a comparison of the galaxy properties between the merger remnants from the CoolStarNoAGN, CoolStarThmAGN, and CoolStarKinAGN simulations, as well as the properties of observed galaxies. The properties of the initial galaxy are also included for reference (shown in cyan). From Panels (a) to (f), we present the $M_{\rm BH}$--$\sigma_\star$ relation, the galaxy size--stellar mass relation, the hot gas X-ray luminosity, the stellar metallicity--stellar mass relation, the stellar mass--halo mass relation, and the dark matter mass fractions within the projected stellar effective radius.

By comparing the properties of our initial galaxies with the properties of the observed galaxies, we find that both our disc and elliptical progenitor galaxies match the observations well. It is worth noting that in Panel (c), which computes the soft X-ray luminosity (0.3--5 keV) emitted by the hot gas, the initial disc galaxy is not shown as it only has a cold gas disc, resulting in an X-ray luminosity value of zero.

We first study the merger remnants from the DD-11-G5 runs, which are represented by filled circles in all panels. In the absence of AGN feedback, the CoolStarNoAGN run exhibits an excessive level of star formation in the galaxy centre, which leads to an elevated $\sigma_{\star}$, the line-of-sight stellar velocity dispersion within the projected half stellar mass radius. Combined with the lack of SMBH accretion, this consequently leads to the deviation from the best-fitting $M_{\rm BH}$--$\sigma_{\star}$ relation shown in Panel (a). Conversely, the merger remnants from both the CoolStarThmAGN and CoolStarKinAGN runs align more closely with the best-fitting $M_{\rm BH}$--$\sigma_{\star}$ relation. Notably, the CoolStarKinAGN exhibits the lowest $\sigma_{\star}$, as a result of the lowest level of star formation among the three simulations. In Panel (b), the CoolStarNoAGN merger remnant shows the most compact stellar size among all runs, primarily due to the steepest stellar density distribution resulting from the intense star formation at the centre. In contrast, the CoolStarKinAGN merger remnant displays the largest stellar size. In Panel (c), the hot gas X-ray luminosity of the CoolStarKinAGN remnant is approximately one order of magnitude lower than the other two runs. This is attributed to the enhanced efficiency of kinetic AGN feedback in expelling the gas from the centre, resulting in reduced central gas density and a more extended hot gas halo \citep[see][for similar conclusions]{Choi2014}. From Panel (d) to (f), the three merger remnants exhibit more similar properties. This similarity arises because the stellar properties within the 30 kpc aperture are less affected by the specific distribution in the very central region.

The merger remnants from the EE-11-G5 runs are plotted with filled diamonds. Unlike the aforementioned DD-11-G5 runs, the stellar properties from all three EE-11-G5 runs are remarkably similar. This similarity comes from the lower overall star formation in elliptical mergers, resulting in negligible differences among the different runs. The most prominent difference among the three runs is found in the hot gas X-ray luminosity, as shown in Panel (c). Specifically, the X-ray luminosity of the CoolStarKinAGN remnant is nearly two orders of magnitude lower than that of the CoolStarThmAGN remnant, further indicating that kinetic AGN feedback is more effective in redistributing the gas from the galaxy centre and thus reducing X-ray luminosity \citep[see also][]{Choi2015}.

It is evident from Fig.~\ref{fig:gal_prop} that for both the DD-11-G5 and EE-11-G5 mergers, the runs with AGN feedback included (i.e. both CoolStarThmAGN and CoolStarKinAGN) generate merger remnants which closely resemble the observed galaxy properties. In Panel (b), some galaxy remnants (e.g. the DD-11-G5 ones) tend to lie below the best-fitting observed relations. This behaviour is anticipated in idealized merger simulations, as the galaxy remnants in our study have undergone a single major merger event. In contrast, real galaxies in the Universe often experience additional minor mergers, which can contribute to an increase in their sizes \citep{Naab2009,Johansson2012}. In Panel (d), the galaxy remnants tend to slightly exceed the observed median stellar metallicity--stellar mass relation. This can be attributed in part to the initial galaxy setup and is further explained by the absence of fresh gas inflow from the intergalactic medium in idealized merger simulations, which would otherwise dilute the metallicity of the interstellar medium and affect the stars formed. In Panel (e), our galaxy remnants tend to be positioned above the stellar mass--halo mass relations \citep[e.g.][]{Moster2013,Kravtsov2018}, which again partially originates from the initial galaxy setup, but the remnants still fall within the range of two times the intrinsic scatter.

We can conclude that our simulations with AGN feedback reproduce reasonably well the observed galaxy properties. The agreement between the simulations and observations demonstrates that the SMBH subgrid model developed in \citetalias{Liao2023} for disc galaxy mergers can be applied to elliptical galaxy mergers as well.

\section{SMBH orbital decay}\label{sec:bh_orbit_decay}

In Section~\ref{sec:gal_prop}, we demonstrated that simulations incorporating different physical processes (e.g. with versus without SMBH accretion and AGN feedback) or employing different numerical implementations (e.g. thermal versus kinetic AGN feedback) yield different galaxy properties, including different central stellar properties, gas distributions, and SMBH masses. Such variations in galaxy properties are expected to result in different rates of orbital decay for the merging SMBHs, influenced by distinct dynamical friction and gravitational potential environments. In this section, we quantitatively compare the SMBH orbital decays across different simulation runs.

Following \citet{Merritt2013}, in the subsequent analyses, we define the SMBH influence radius as the radius where the circular velocity around an SMBH equals the one-dimensional (or line-of-sight) stellar velocity dispersion,
\begin{equation}
    R_{\rm infl} = \frac{G M_{\rm BH}}{\sigma_\star^2}.
\end{equation}
When the separation of two SMBHs reaches roughly the influence radius of the more massive SMBH, $\Delta d_{\rm BH} \sim R_{\rm infl}$, the two SMBHs form a bound binary. Usually, this is regarded as the end of the dynamical friction-dominated phase, after which the interactions between star particles and the SMBH binary (i.e. three-body interactions) start to play a role in the SMBH orbital decay. 

The hard binary separation is defined as \citep{Merritt2013}
\begin{equation}
    R_{\rm hard} = \frac{G M_{\rm red}}{4 \sigma_\star^2} = \frac{M_{\rm BH, 2}}{M_{\rm bin}}\frac{R_{\rm infl}}{4},
\end{equation}
where $M_{\rm red} = M_{\rm BH,1} M_{\rm BH,2} / M_{\rm bin}$ is the reduced mass. The hard binary separation is roughly the radius where the specific binding energy of the SMBH binary equals the specific kinetic energy of the surrounding stars. When the semimajor axis of the SMBH binary reaches the hard binary separation, $a \sim R_{\rm hard}$, the SMBH orbital decay undergoes a complete transition from dynamical friction to being entirely dominated by three-body interactions. In our simulations, typically $R_{\rm infl}$ are ${\sim} 10$ pc while $R_{\rm hard}$ are ${\sim} 1$ pc.

\subsection{Overall evolution of SMBHs and their embedded galaxy centres}

\begin{figure*} 
\centering\includegraphics[width=400pt]{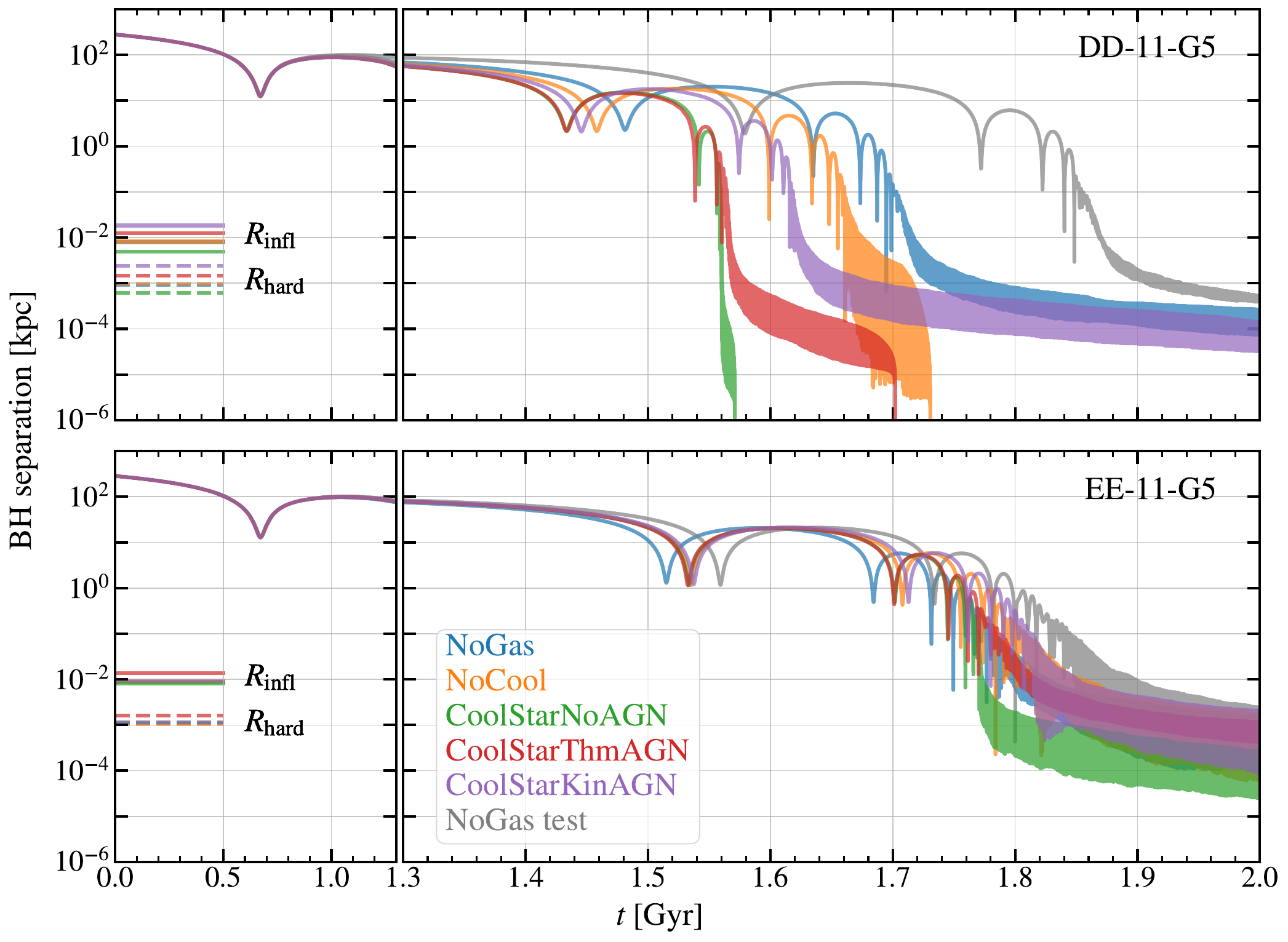}
\caption{Time evolution of the SMBH separation for both the disc (top) and the elliptical (bottom) galaxy mergers. The NoGas, NoCool, CoolStarNoAGN, CoolStarThmAGN, and CoolStarKinAGN runs are plotted with blue, orange, green, red, and purple lines, respectively. We also plot the additional NoGas test runs which exclude the gas particles in the initial condition (grey) for comparison. The horizontal line segments in the left panels mark the influence radius $R_{\rm infl}$ (solid) and the hard binary separation $R_{\rm hard}$ (dashed).}
\label{fig:bh_sep}
\end{figure*}

\begin{figure*} 
\centering\includegraphics[width=400pt]{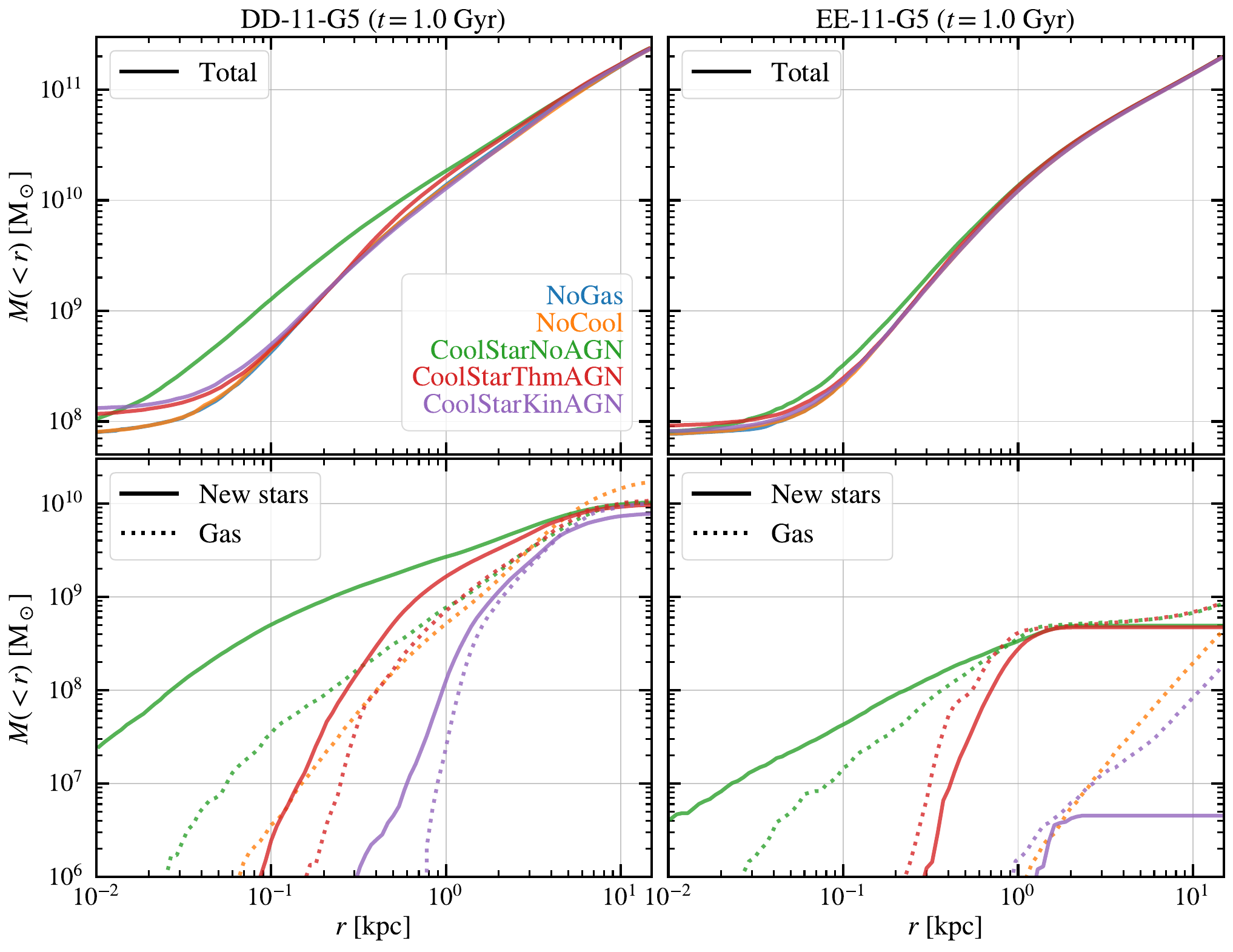}
\caption{Enclosed mass profiles centring on one of the SMBHs in DD-11-G5 (left) and EE-11-G5 (right) runs at the first apocentre ($t \approx 1$ Gyr). In this comparison, the SMBHs from different simulations are matched using their respective particle IDs. The upper panels show the enclosed total mass (including SMBH, stars, gas, and dark matter) profiles, whereas the lower panels plot the enclosed mass profiles for new stars (solid) and gas (dotted). Here the new stars are those formed after the start of the simulation. The colour convention used to denote the different simulations is given in the top-left panel. Note that the profiles centred on the other SMBH are very similar to the profiles plotted here.}
\label{fig:enclose_mass_1Gyr}
\end{figure*}

\begin{figure*} 
\centering\includegraphics[width=400pt]{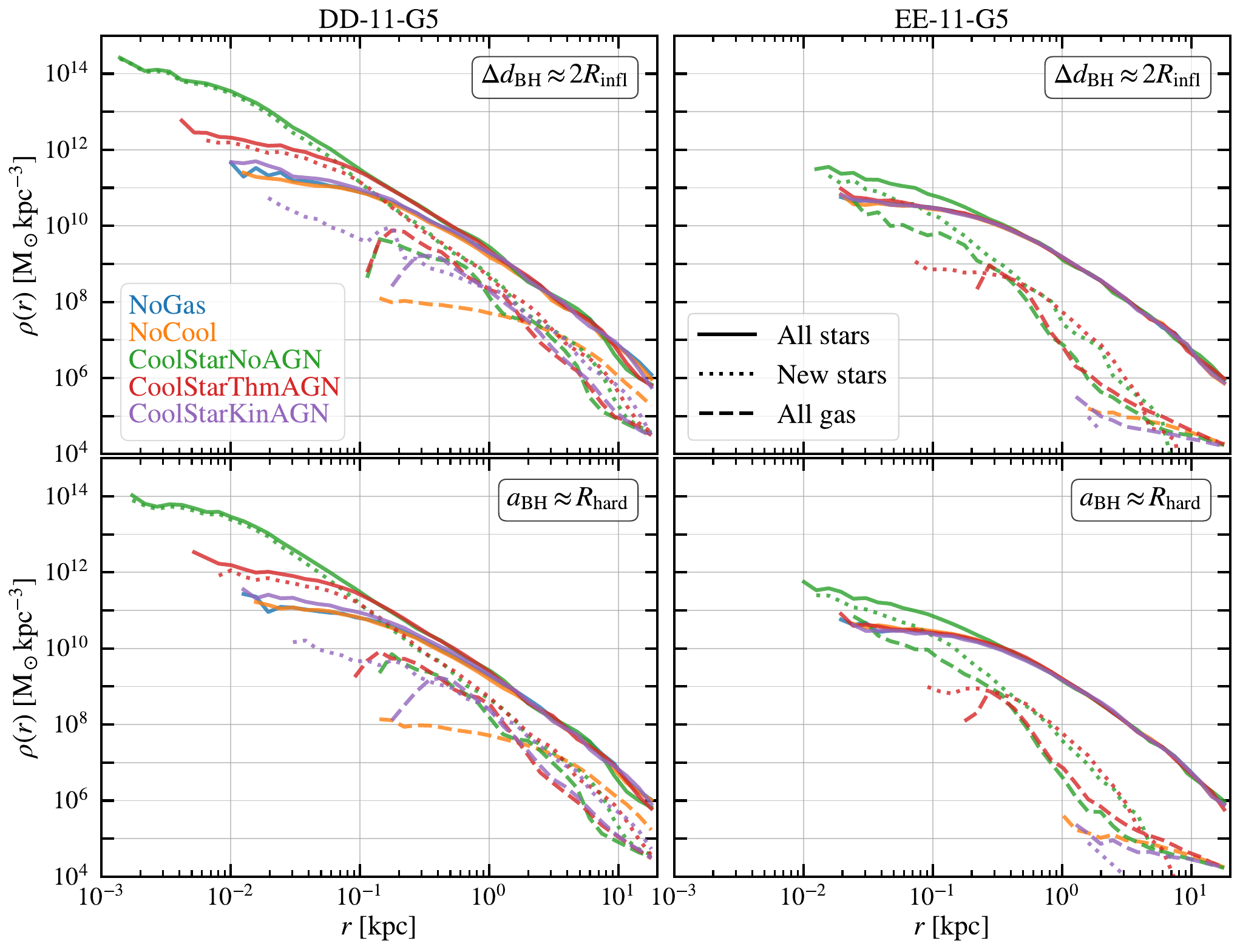}
\caption{Spherically averaged density profiles centred on the CoM position of the two SMBHs for the galaxy merger remnants in the DD-11-G5 (left) and the EE-11-G5 (right) runs. The density profiles computed from all star particles, new star particles only, and all gas particles are plotted with solid, dotted, and dashed curves, respectively. Only bins with at least ten particles are shown. Different simulations are distinguished by colours according to the convention outlined in the top-left panel. The upper panels display the density profiles when the SMBH separation is roughly twice the SMBH influence radius (i.e. the SMBH orbital decay is dominated by the dynamical friction from surrounding stars and gas). The lower panels show the density profiles when the semimajor axis of the SMBH binary reaches approximately the hard separation (i.e. the onset of the SMBH binary hardening phase which is dominated by the three-body interactions between the SMBH binary and the individual stars).}
\label{fig:rho_df_rhard}
\end{figure*}

In Fig.~\ref{fig:bh_sep}, we present the time-evolution of the SMBH separation in the first 2 Gyr for all simulation runs. In the left panels, we mark the SMBH influence radius, $R_{\rm infl}$, and the hard binary separation, $R_{\rm hard}$, from different runs.

We first focus on the DD-11-G5 mergers, which are summarized in the top panels. The evolution of the different runs during the first ${\sim} 1$ Gyr are quite similar as the plotted lines are almost on top of each other. As discussed in Section~\ref{subsec:overview}, initially the two SMBHs and the bound galaxy centres that they are embedded in move together following the input parabolic orbits until they experience strong tidal interactions. During the first pericentre passage, the tidal effects are not identical for galaxies in different runs as they have distinct galaxy compositions (i.e. NoGas versus other runs) and they have experienced distinct star formation and SMBH growth histories (i.e. runs except NoGas and NoCool; see Fig.~\ref{fig:sfr_bhar}) which lead to different central galaxy properties. As a result, the time-evolution of the SMBH separation in different runs start to deviate from each other and the differences become more noticeable after the first apocentre (e.g. at $t \sim 1$ Gyr). After the first apocentre, the NoGas run exhibits the slowest orbital decay, followed by the NoCool run, the CoolStarKinAGN run, the CoolStarThmAGN run, and finally the CoolStarNoAGN run which shows the most rapid orbital decay.

The comparison between the NoGas and the NoCool runs showcases the impact of gas, i.e. the gas pressure assists in the conversion of orbital energy and angular momentum into internal motions, resulting in a more rapid SMBH orbital decay. Similar phenomenon was observed in \citet{Barnes1996} for their collisionless and gaseous paired runs of a retrograde disc galaxy merger.

Compared to the NoCool run, the runs with gas cooling and star formation (i.e. CoolStarKinAGN, CoolStarThmAGN, and CoolStarNoAGN) display even more rapid orbital decay, which is an outcome of the more bound galaxy centres. The enclosed total mass profiles (including SMBH, star, gas, and dark matter) at $t = 1$ Gyr from different runs are shown in the top-left panel of Fig.~\ref{fig:enclose_mass_1Gyr}. We can easily see that the central matter distributions tend to be more concentrated in those runs including gas cooling and star formation. The enclosed mass profiles for the new stars and gas are plotted in the bottom-left panel of Fig.~\ref{fig:enclose_mass_1Gyr}. Here, the new stars are defined as the stars formed after the start of the simulation, i.e. stars with a formation time $t_{\rm SF} \geq t_0$. The CoolStarNoAGN run has a significant amount of new stars formed in the very inner region (i.e. $r \la 1$ kpc), while the AGN feedback suppresses the star formation in the very inner region in the CoolStarThmAGN and the CoolStarKinAGN runs. Compared to the NoCool run, the gas distributions in the CoolStarThmAGN and the CoolStarKinAGN runs are less concentrated due to two factors, i.e. some inner gas particles have already been converted into stars and more importantly the AGN feedback has heated and pushed the gas to the outer region.

The more concentrated galaxy centres are more bound, and thus they can resist the tidal stripping effect more effectively and sustain higher masses for a longer time. As a result, they experience stronger dynamical friction and their SMBH orbits decay faster. This explains why the CoolStarNoAGN run exhibits the most rapid SMBH orbital decay, followed by the CoolStarThmAGN and the CoolStarKinAGN runs, and why these three runs show more rapid decays than the NoCool and the NoGas runs. This picture is in line with the work of \citet{Callegari2009}, who found that in minor mergers of disc galaxies, incorporating radiative cooling and star formation creates more bound satellite galaxy centres and consequently enhances the SMBH pair formation, in contrast to dry merger simulations without gas. The recent work of \citet{Chen2023} has also suggested the important role of high central stellar density in assisting the orbital decay of SMBHs.

The EE-11-G5 mergers are summarized in the bottom panels of Fig.~\ref{fig:bh_sep}. Similar to the DD-11-G5 case, during the first ${\sim} 1$ Gyr, the evolution of SMBH separations from different EE-11-G5 runs are almost identical, since the SMBHs and their embedded galaxy centres are moving together and the bound masses are very similar across all runs. In addition, we notice that the evolution during the first ${\sim} 1$ Gyr are fairly similar between the DD-11-G5 and the EE-11-G5 runs, as a result of the designed fraternal twin initial conditions.

After the first apocentre, we start to observe some noticeable differences among the different runs. However, overall the differences among the different runs are smaller than those in the DD-11-G5 runs, because the star formation and SMBH growth are much weaker in the EE-11-G5 runs and the enclosed mass profiles are more similar for the different runs (see the right panels of Fig.~\ref{fig:enclose_mass_1Gyr}). 

In the EE-11-G5 runs, the largest difference in the SMBH separation evolution is between the NoGas run and the other runs that include gas. Interestingly, in contrast to the DD-11-G5 case, here the SMBHs in the NoGas run reaches their second pericentre passages earlier than the runs including gas. This is caused by the fact that we convert all gas particles in the hot gas halo into star particles in the initial condition for the NoGas run, and the evolution of this `stellar halo' is distinct from that of the gas halo, creating different gravitational drags on the galaxy centres. We have performed a test run by simply removing the gas particles in the hot gas halo to create an initial condition, which is dubbed `NoGas test' and is plotted with grey in the bottom panel of Fig.~\ref{fig:bh_sep}. We can see that without this `stellar halo', the SMBHs reach their second pericentre passage later compared to the runs including gas, in line with the DD-11-G5 case. For the sake of completeness, we also plot the `NoGas test' run for the DD-11-G5 merger in the top panel of Fig.~\ref{fig:bh_sep}, illustrating how the least concentrated galaxy centre results in the slowest orbital decay rate.

Among all EE-11-G5 runs, the CoolStarNoAGN run has the most star formation in the galaxy centre, creating the most bound central regions and consequently having its SMBH separation reaching $R_{\rm infl}$ first (i.e. exhibiting the most rapid orbital decay). In Section~\ref{subsec:dyn_fri}, we will show that the gas dynamical friction also contributes to the SMBH orbital decay in the CoolStarNoAGN case. The CoolStarThmAGN run has the second most rapid orbital decay, followed by the NoCool and the CoolStarKinAGN runs. As the star formation is completely quenched after the first pericentre passage and kinetic AGN feedback is effective in maintaining a hot gas halo, the CoolStarKinAGN run displays relatively similar SMBH orbital decay as the NoCool run.

Overall, the orbital decay of SMBHs in the EE-11-G5 case is slower than that in the DD-11-G5 case. This is because compared to the DD-11-G5 runs, the EE-11-G5 runs exhibit less star formation and less SMBH growth, resulting in less bound galaxy centres, weaker dynamical friction, and slower SMBH orbital decays.

\subsection{Dynamical friction on SMBHs: stars versus gas}\label{subsec:dyn_fri}

As the separation between the SMBHs decreases, the matter initially bound to the galaxy centre continues to be tidally stripped away. During the last stage of the dynamical friction-dominated phase, most of the matter that was initially bound to the SMBHs has been stripped away, leaving the two SMBHs effectively as isolate massive point particles moving within the stellar and gaseous background.

To address the relative importance of stellar and gaseous dynamical friction, we first look at the stellar (solid lines) and gas (dashed lines) density profiles when the SMBH separations are approximately twice the SMBH influence radius (i.e. $\Delta d_{\rm BH} \sim 20$ pc). These density profiles are presented in the top panels of Fig.~\ref{fig:rho_df_rhard}. Note that here the profiles are centred on the CoM position of the two SMBHs. We can clearly see that in all simulations including gas, the inner mass distributions are dominated by stars. In both the DD-11-G5 and EE-11-G5 cases with AGN feedback, the SMBHs push gas away, leading to a significant reduction in the gas density within ${\sim} 100$ pc. The kinetic AGN feedback in the EE-11-G5 case is particularly effective in achieving this. Without AGN feedback, one would expect gas to cool down and condense in the galaxy centre, resulting in a higher gas central density, as observed in the EE-11-G5 CoolStarNoAGN run. However, in DD-11-G5 CoolStarNoAGN, the unrealistically massive stellar centre produces strong supernova feedback which pushes the gas to larger radii, resulting in a gas density profile similar to the CoolStarThmAGN one.

The central density has an important impact on the dynamical friction. To further provide a quantitative comparison, we estimate the stellar and gaseous dynamical frictions from our simulations. The stellar dynamical friction can be modelled as \citep[][]{Chandrasekhar1943,Binney2008}
\begin{equation}
    F_{\rm df, \star} = - \frac{4 \pi G^2 M_{\rm BH}^2 \rho_{\star}}{v_{\rm rel, \star}^2} I_{\star},
\end{equation}
where $\rho_{\star}$ is the central stellar density, $v_{\rm rel, \star}$ is the magnitude of the relative velocity between the SMBH and the stellar background, and 
\begin{equation}
    I_{\star} = \ln \Lambda \left[{\rm erf}(X) - \frac{2X}{\sqrt{\pi}}e^{-X^2}\right].
\end{equation}
Here, $\ln \Lambda$ is the Coulomb logarithm, $X = v_{\rm rel, \star}/(\sqrt{2}\sigma_{\star})$ is the ratio between the SMBH--star relative velocity and the stellar velocity dispersion, and ${\rm erf}(X)$ is the error function. On the other hand, the gas dynamical friction can be estimated by \citep{Ostriker1999}
\begin{equation}
    F_{\rm df, gas} = - \frac{4 \pi G^2 M_{\rm BH}^2 \rho_{\rm gas}}{v_{\rm rel, gas}^2} I_{\rm gas},
\end{equation}
where $\rho_{\rm gas}$ is the central gas density, $v_{\rm rel, gas}$ is the magnitude of the relative velocity between the SMBH and the gas background, and the velocity-dependent factor, $I_{\rm gas}$, has different forms in the subsonic and the supersonic regimes, i.e. 
\begin{equation}
    I_{\rm gas} = \frac{1}{2}\ln \left(\frac{1 + \mathcal{M}}{1 - \mathcal{M}}\right) - \mathcal{M}, ~ ({\rm for} ~ \mathcal{M} < 1)
\end{equation}
\begin{equation}
    I_{\rm gas} = \frac{1}{2} \ln \left(1 - \frac{1}{\mathcal{M}^2}\right) + \ln \Lambda. ~ ({\rm for} ~ \mathcal{M} > 1)
\end{equation}
Here, $\mathcal{M} = v_{\rm rel, gas} / c_{\rm s}$ is the Mach number. Therefore, the ratio between the stellar and the gaseous dynamical frictions is
\begin{equation}
    \frac{F_{\rm df, \star}}{F_{\rm df, gas}} = \frac{\rho_{\star}}{\rho_{\rm gas}} \left(\frac{v_{\rm rel, gas}}{v_{\rm rel, \star}}\right)^2 \frac{I_{\star}}{I_{\rm gas}}.
\end{equation}
It was shown in \citet{Ostriker1999} that, if $\rho_{\rm gas} = \rho_{\star}$, $v_{\rm rel, gas} = v_{\rm rel, \star}$, and $c_{\rm s} = \sigma_{\star}$, then, in the supersonic (subsonic) regime, the gaseous dynamical friction is stronger (weaker) than the stellar one.

To estimate the force ratio, in our calculations, we use the star particles within the radius\footnote{Note that the specific choice of this radius has a quantitative impact on the computed values of the stellar properties and the force ratio. However, we have performed tests and found that the conclusions presented in this subsection remained unchanged for other choices of radius, such as 50 pc, 500 pc, and 1 kpc.} of $100$ pc centring on the SMBH to compute the related stellar quantifies, i.e. $\rho_{\star}$ is estimated as the mean stellar density within $100$ pc, $v_{\rm rel, \star}$ is the relative velocity between the SMBH and the CoM velocity of the used star particles, and $\sigma_{\star}$ is their velocity dispersion. The gas related quantities, $\rho_{\rm gas}$, $v_{\rm rel, gas}$, and $c_{\rm s}$, are estimated at the SMBH position using the SPH approach. The Coulomb logarithm is estimated as \citep[][]{Binney2008}
\begin{equation}
    \ln \Lambda = \ln \left(\frac{b_{\rm max}}{b_{\rm min}}\right) \approx \ln \left(\frac{R_{200}}{G M_{\rm BH} / V_{200}^2}\right) \approx 10,
\end{equation}
where $V_{200}^2 = G M_{200} / R_{200}$ is the squared virial velocity, and we have adopted $M_{200} \approx 3 \times 10^{12}~{\rm M}_{\sun}$ and $M_{\rm BH} \approx 10^{8}~{\rm M}_{\sun}$. The detailed results are summarized in Table~\ref{tab:dyn_fri}.

\begin{table*}
\begin{threeparttable}
\caption{Comparison between stellar and gaseous dynamical friction. As the results for the two SMBHs are very similar, here we only present the computed properties of one SMBH. The information is derived from the snapshots shown in the top panels of Fig.~\ref{fig:rho_df_rhard} (i.e. when $\Delta d_{\rm BH} \approx 2 R_{\rm infl}$). From the left, the galaxy merger name, the simulation setup, the gas density at the SMBH position, the SMBH--gas relative velocity, the sound speed at the SMBH position, the Mach number, the mean stellar density within 100 pc, the relative velocity between the SMBH and the CoM velocity of star particles within 100 pc, the stellar velocity dispersion computed from star particles within 100 pc, and the ratio between the estimated stellar and gaseous dynamical friction forces.}
\label{tab:dyn_fri}
\begin{tabular}{llcccccccccccc}
\hline
Galaxy & Simulation & $\rho_{\rm gas}$ & $v_{\rm rel, gas}$ & $c_{\rm s}$ & $\mathcal{M}$ & $\rho_{\star}$ & $v_{\rm rel, \star}$ & $\sigma_{\star}$ & $F_{\rm df,\star}/F_{\rm df, gas}$ \\
merger & & $[{\rm M}_{\sun}{\rm kpc}^{-3}]$ & [km s$^{-1}$] & [km s$^{-1}$] & & $[{\rm M}_{\sun}{\rm kpc}^{-3}]$ & [km s$^{-1}$] & [km s$^{-1}$] & \\
\hline
DD-11-G5 & CoolStarNoAGN & $3.7 \times 10^{9}$ & $276.8$ & $656.2$ & $0.42$ & $3.6 \times 10^{11}$ & $607.8$ & $306.7$ & $5.3 \times 10^3$\\
DD-11-G5 & CoolStarThmAGN & $1.9 \times 10^{9}$ & $350.5$ & $980.5$ & $0.36$ & $1.6 \times 10^{11}$ & $100.6$ & $269.1$ & $8.2 \times 10^3$\\
DD-11-G5 & CoolStarKinAGN & $1.7 \times 10^{8}$ & $606.7$ & $353.3$ & $1.72$ & $4.7 \times 10^{10}$ & $213.3$ & $221.3$ & $4.2 \times 10^2$\\
EE-11-G5 & CoolStarNoAGN & $9.9 \times 10^{10}$ & $52.6$ & $13.4$ & $3.91$ & $3.5 \times 10^{10}$ & $49.6$ & $173.7$ & $2.4 \times 10^{-3}$\\
EE-11-G5 & CoolStarThmAGN & $7.1 \times 10^{7}$ & $226.7$ & $540.6$ & $0.42$ & $1.3 \times 10^{10}$ & $155.6$ & $190.4$ & $1.7 \times 10^{4}$\\
EE-11-G5 & CoolStarKinAGN & $1.2 \times 10^{5}$ & $160.6$ & $602.4$ & $0.27$ & $1.2 \times 10^{10}$ & $80.8$ & $182.6$ & $1.2 \times 10^{7}$\\
\hline
\end{tabular}
\end{threeparttable}
\end{table*}

In most runs, except for DD-11-G5 CoolStarKinAGN and EE-11-G5 CoolStarNoAGN, the gas density is significantly lower than the stellar density and the SMBH motion is subsonic (i.e. $\mathcal{M} \sim 0.4$), therefore the stellar dynamical friction dominates over the gaseous one. In DD-11-G5 CoolStarKinAGN, even though the SMBH moves in the supersonic regime ($\mathcal{M} \sim 1.7$), the stellar density is much higher than the gas density, leading to the dominance of stellar dynamical friction. Interestingly, in EE-11-G5 CoolStarNoAGN, due to weak stellar feedback and the absence of AGN feedback, the central gas density becomes comparable to the stellar density. Furthermore, the SMBH moves in the supersonic regime, resulting in the dominance of gas dynamical friction. Note that we have performed a similar analysis in later snapshots (with $\Delta d_{\rm BH} \sim 5$--$10$ pc), and the results are qualitatively similar.

From these comparisons, we can conclude that when AGN feedback is included, the gas density around the SMBHs is drastically reduced, resulting in a decrease in gas dynamical friction on the SMBHs. Consequently, stellar dynamical friction becomes the dominant force affecting the SMBH motion. Importantly, this conclusion is applicable to both the thermal and kinetic AGN feedback implementations. 

Our conclusions echo the previous works of e.g. \citet{Mayer2007} and \citet{SouzaLima2017}. \citet{Mayer2007} demonstrated that in a coplanar and prograde merger of two equal-mass disc galaxies, gas funnelled by tidal torques forms a massive nuclear disc in the remnant centre, leading to SMBH orbital decay driven primarily by gas dynamical friction instead of stellar dynamical friction. We attribute the main cause of this to the lack of AGN feedback in their simulations, resulting in excessive gas accumulation in the remnant centre, and the lack of feedback from supernova explosions which has further exacerbated this effect. In contrast, \citet{SouzaLima2017} showed that once AGN feedback is included, the spherically distributed feedback can generate a hot bubble around the SMBH and consequently shuts off gas dynamical friction, a phenomenon named `wake evacuation' by the authors. Recently, \citet{Bollati2023} also concluded from their simulations that the inclusion of radiative AGN feedback can lead to inefficient or even reversed gas dynamical friction, delaying the orbital decay of SMBH pairs compared to the no feedback case. Our results also agree with those of \citet{Pfister2017}, who found that dynamical friction from stars predominantly drives the orbital decay of SMBHs in their galaxy merger simulations incorporating AGN feedback, whilst the contribution from gas is negligible.

It is worth emphasizing that our comparison of stellar and gaseous dynamical frictions above is based on the dynamical friction models from \citet{Binney2008} and \citet{Ostriker1999}. In our simulations, the dynamical friction acting on SMBHs from stars is accurately accounted for, thanks to the precise computation of non-softened gravity between the SMBHs and star particles using a regularized approach \citep{Karl2015,Rantala2017}. In contrast, the gravity between the SMBHs and gas is softened in the {\sc ketju} code, which can potentially lead to the underestimation of gas dynamical friction since the gas particles with impact factors lower than $2.8$ times the softening length are not properly modelled. However, this does not affect our conclusion that the stellar dynamical friction dominates over the gaseous one in our runs with AGN feedback, as AGN feedback plays a more significant role in affecting the gas density and thus the gas friction. In addition, we note that the accretion of gas mass and momentum into SMBHs in CoolStarThmAGN and CoolStarKinAGN runs also contribute to the SMBH orbital decay.

Approximately when the separation of the two SMBHs reaches the influence radius of the more massive SMBH, they form a bound binary. As can be seen from Fig.~\ref{fig:bh_sep}, in all runs, the SMBH separations drop very rapidly in the regime of $R_{\rm hard} \la \Delta d_{\rm BH} \la R_{\rm infl}$, as a result of the combined effects from dynamical friction and the gravitational slingshot interactions between stars and the SMBH binary. When the separation (or the binary semimajor axis) reaches ${\sim} R_{\rm hard}$, a hard SMBH binary forms and its subsequent orbital decay is discussed in the following subsection.

\subsection{SMBH binary hardening and GW emission phases}\label{subsec:bh_hard}

\begin{figure*} 
\centering\includegraphics[width=400pt]{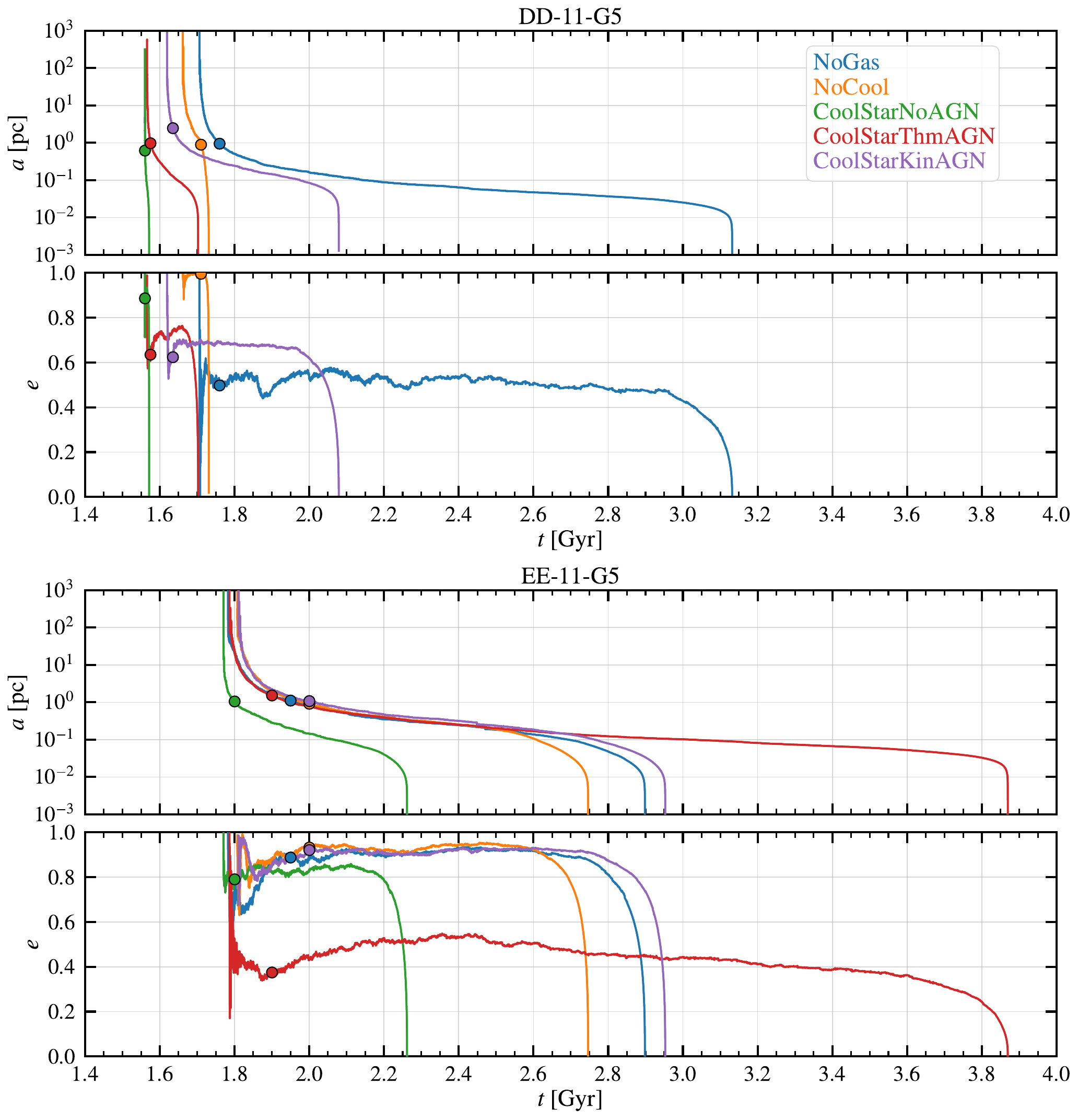}
\caption{Time-evolution of the PN-corrected orbital parameters, semimajor axis $a$ and eccentricity $e$, in the DD-11-G5 (upper) and the EE-11-G5 (lower) runs. The filled circles mark the time when the SMBH binaries become hard. Different colours are adopted to distinguish different simulations as given in the upper semimajor axis panel.}
\label{fig:orb_params}
\end{figure*}

\begin{figure*} 
\centering\includegraphics[width=425pt]{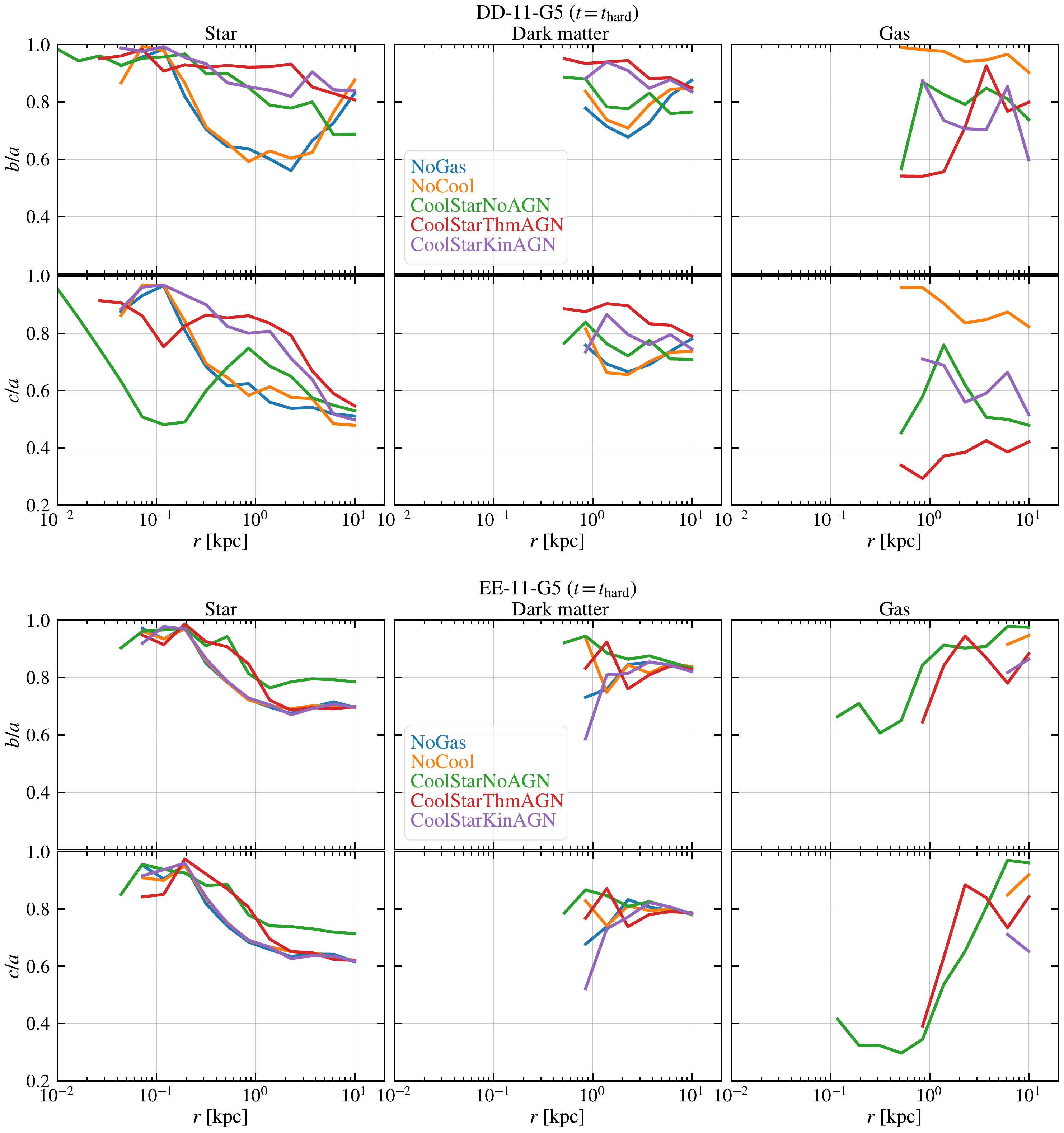}
\caption{Axis ratios of the galaxy merger remnants as a function of distance from the CoM of the SMBH binary at $t = t_{\rm hard}$. The upper and lower panels show the results from the DD-11-G5 and EE-11-G5 simulations, respectively. From left to right, the shape profiles of the stellar, dark matter, and gas components are plotted. For each component, the top and bottom panels show the intermediate-to-major axis ratios ($b/a$) and the minor-to-major axis ratios ($c/a$), respectively. To ensure a robust estimation of the shapes, we plot the profiles only for radii where the enclosed particle number exceeds 300 \citep{Bett2007}. The merger remnants in all runs exhibit triaxiality in the different components all the way to the centre. The inclusion of gas cooling, stellar physical processes, and AGN feedback has a significant impact on the triaxiality of the matter distribution, especially in disc galaxy mergers and for the gas component.}
\label{fig:axis_ratio}
\end{figure*}

In Fig.~\ref{fig:orb_params}, we plot the time-evolution of the PN-corrected orbital parameters \citep{Memmesheimer2004,Mannerkoski2019}, specifically the semimajor axis $a$ and the eccentricity $e$, of the SMBH binaries from the different runs. The filled circles mark the time $t = t_{\rm hard}$ when a binary becomes hard, i.e. when the semimajor axis of an SMBH binary reaches approximately the hard binary separation $R_{\rm hard} \sim 1$ pc.

After $t_{\rm hard}$, during the gravitational slingshot interaction phase, the central stellar properties have an important impact on driving the binary hardening \citep[e.g.][]{Quinlan1996,Quinlan1997}. The density profiles centred on the CoM position of the SMBH binary are plotted in the bottom panels of Fig.~\ref{fig:rho_df_rhard}. For both disc and elliptical galaxy mergers, the central stellar densities in the CoolStarNoAGN runs are significantly higher than those in other runs, therefore, it is unsurprising to find that the SMBH binaries in both the DD-11-G5 and EE-11-G5 CoolStarNoAGN runs exhibit the shortest merger time-scales. Especially, for the DD-11-G5 CoolStarNoAGN case, the SMBH merger time-scale is as short as ${\sim} 10$ Myr.

The departure from the spherical potential can also lead to rapid binary hardening, which has been regarded as a possible solution to the final parsec problem \citep{Milosavljevic2003}, i.e. stellar orbits in a triaxial potential can refill the loss cone more efficiently compared to the case of a spherical potential, resulting in efficient hardening of SMBH binaries \citep{Merritt2004,Berczik2006,Khan2011,Preto2011}. In Fig.~\ref{fig:axis_ratio}, we show the axis ratio (including the intermediate-to-major axis ratio, $b/a$, and the minor-to-major axis ratio, $c/a$) profiles of different components for our galaxy remnants at $t_{\rm hard}$. The axis ratios are computed following the iterative method described in \citet{Katz1991}. For a given radius of $r$, we first use all particles within $r$ to compute the moment of inertia tensor and obtain the axis ratios from the tensor's eigenvalues. Then, new axis ratios are computed using only the particles enclosed within the ellipsoidal volume with the previously determined axis ratios. This procedure is repeated for the ellipsoidal volume with the new axis ratios, and iterated until the axis ratios converge. 

From Fig.~\ref{fig:axis_ratio}, we can see that the merger remnants in all runs exhibit triaxiality all the way to the centre. The NoGas and NoCool runs show quite similar axis ratio profiles in the stellar and dark matter components. Conversely, in the other runs, the galaxy formation processes (especially AGN feedback) show a significant impact on the triaxial matter distributions, especially in disc galaxy mergers and for the gas component. With weaker AGN feedback, the stellar and gas components tend to show more triaxial distributions at smaller radii, as a result of more efficient cooling and star formation. At larger radii (i.e. $\ga 1$ kpc), compared to the NoGas and NoCool cases, the energetic galaxy formation processes have resulted in more dynamically hot and thus less triaxial distributions in the stellar and dark matter components. Overall, the disc galaxy merger remnants tend to be more triaxial compared to the elliptical ones.

The eccentricity of an SMBH binary when it becomes hard, $e_{\rm hard}$, is another crucial factor affecting the merger time-scale. During the gravitational slingshot interaction phase, the eccentricity only shows marginal evolution, i.e. usually it tends to grow slightly according to the three-body scattering experiments \citep[e.g.][]{Mikkola1992,Quinlan1996,Sesana2006}, which is also observed in our runs. According to \citet{Peters1963} and \citet{Peters1964}, the transition to the GW emission-driven phase and the efficiency of the GW emission depend sensitively on the binary eccentricity. Specifically, during the GW emission-driven phase, the decaying rates in the semimajor axis and the eccentricity can be approximately (at the 2.5PN level) described by \citep{Peters1964}
\begin{equation}
    \frac{{\rm d} a}{{\rm d} t} = - \frac{64}{5} \frac{G^3 M_{\rm BH,1} M_{\rm BH,2} M_{\rm bin}}{c^5 a^3 \left(1 - e^2\right)^{7/2}} \left(1 + \frac{73}{24}e^2 + \frac{37}{96}e^4\right),
\end{equation}
\begin{equation}
    \frac{{\rm d} e}{{\rm d} t} = - \frac{304}{15} \frac{G^3 M_{\rm BH,1} M_{\rm BH,2} M_{\rm bin}}{c^5 a^4 \left(1 - e^2\right)^{5/2}} e \left(1 + \frac{121}{304}e^2\right).
\end{equation}
A higher eccentricity value closer to 1 will significantly accelerate the GW emission phase due to the factor of $(1 - e^2)^{-7/2}$. This is the dominant mechanism in driving the rapid SMBH merger in the DD-11-G5 NoCool run, i.e. the SMBH binary has an $e_{\rm hard} \sim 0.995$ and the GW emission rapidly dominates the orbital decay over other processes, resulting in a very short merger time-scale after $t_{\rm hard}$ (i.e. ${\sim} 20$ Myr). Similarly, the low $e_{\rm hard} \sim 0.5$ is partly responsible for why the NoGas (CoolStarThmAGN) run of the DD-11-G5 (EE-11-G5) merger has a much longer merger times-scale compared to other runs with higher $e_{\rm hard}$.

We note that in gas-free $N$-body simulations, $e_{\rm hard}$ has been found to suffer from stochasticity due to the random encounters with stars \citep[e.g.][]{Quinlan1997,Berczik2005,Nasim2020,Gualandris2022,Rawlings2023}. Specifically, pure $N$-body simulations of galaxy mergers starting from different realizations of initial conditions can lead to distinct $e_{\rm hard}$ for their hard binaries. Previous works have suggested that the scatter of $e_{\rm hard}$ due to this stochasticity can be reduced when the simulations are run with higher mass resolutions \citep[or equivalently more particles; see e.g.][]{Quinlan1997,Berczik2005,Nasim2020}. For the simulations including gas in this study, $e_{\rm hard}$ can have higher degrees of stochasticity compared to the gas-free case, as the galaxy formation subgrid model introduces additional stochastic processes. Another uncertainty in the binary eccentricity comes from the torque interaction between the binary and its circumbinary disc. Recently, some small-scale hydrodynamical simulations \citep[e.g.][]{DOrazio2021,Zrake2021,Siwek2023} suggested that the binary--disc torque interaction might drive the eccentricity toward two `attractor' solutions, i.e. an eccentric and equal-mass binary with $e \ga 0.1$ ($e \la 0.1$) evolves to an equilibrium value of $e_{\rm eq} \sim 0.5$ ($e_{\rm eq} \sim 0$). 

Although the detailed SMBH merger time-scales of the different simulations are affected by varying $e_{\rm hard}$, in Fig.~\ref{fig:orb_params}, we can still see an overall trend that the SMBH binaries in disc galaxy mergers tend to merge more rapidly than those in elliptical mergers. Specifically, the SMBHs in the DD-11-G5 runs tend to cluster around a lower $e_{\rm hard} \sim 0.6$ and merge within a few hundred Myr (i.e. $\la 500$ Myr) after they become hard (except the NoGas run which has a low central stellar density and a low $e_{\rm hard}$). In contrast, the SMBHs in the EE-11-G5 runs tend to cluster around a higher $e_{\rm hard} \sim 0.9$ but merge at time-scales $\ga 1$ Gyr after $t_{\rm hard}$ (except the CoolStarNoAGN run which has the highest central stellar density among all elliptical merger runs). In Appendix~\ref{ap:res_study}, we demonstrate that the contrast in SMBH merger time-scales between disc and elliptical galaxy mergers remains consistent across runs with different mass resolutions. Such a difference in SMBH merger time-scales aligns with the higher central stellar densities and more triaxial potentials observed in disc galaxy merger remnants compared to elliptical cases.

For a direct comparison of SMBH merger time-scales across runs including different physical processes, it is necessary to fix $e_{\rm hard}$ to similar values to mitigate the significant impact from varying eccentricities. In Paper I, we rerun the simulations from $t_{\rm hard}$ by resetting $e_{\rm hard}$ to the same value, which allow us to perform a more direct and more quantitative comparison of the merger time-scales. In Paper I, we demonstrate that the torque interaction between the binary and the circumbinary disc only plays a minimal role in affecting the shrinking of the orbit in the G5 galaxy merger.

\section{Conclusions}\label{sec:con}

In this work, we presented the RABBITS simulations and investigated how the presence and the specific implementation (i.e. thermal versus kinetic) of AGN feedback influence the properties of merging galaxies and the orbital decay of their SMBHs. We considered both disc-disc and elliptical-elliptical galaxy mergers, and performed systematically controlled simulation sets including different physical processes (i.e. NoGas, NoCool, CoolStarNoAGN, CoolStarThmAGN, and CoolStarKinAGN; see Table~\ref{tab:sim_set} for details).

Our key findings are summarized as follows.
\begin{description}
\item {\bf AGN feedback plays a critical role in shaping the galaxy properties:}
\begin{enumerate}
\item Simulations with cooling and stellar processes (i.e. star formation and stellar feedback), but without AGN feedback, exhibit excessive star formation in the galaxy centre, resulting in galaxy remnants which deviate from some observed galaxy scaling relations (e.g. the $M_{\rm BH}$--$\sigma_{\star}$ relation). 
\item Compared to thermal AGN feedback, kinetic AGN feedback is more effective in removing gas from the galaxy centre, preventing cooling flows, quenching star formation, and maintaining a red and quiescent galaxy remnant. 
\end{enumerate}

\item {\bf Different central galaxy properties caused by different AGN models further affect the SMBH orbital decaying process:}
\begin{enumerate}\setcounter{enumi}{2}
\item Galaxies with higher star formation and higher SMBH masses (attributed to increased accretion) tend to possess denser central regions which are more resistant to tidal stripping, experience larger dynamical friction, and consequently form bound SMBH binaries earlier.
\item In the dynamical friction phase, as AGN feedback effectively reduces the gas density around SMBHs, stellar dynamical friction dominates over the gaseous friction in shrinking the SMBH orbit.
\item In the binary hardening phase, compared to elliptical-elliptical galaxy mergers, disc-disc mergers tend to have more rapid SMBH hardening and shorter SMBH merger time-scales (i.e. $\la 500$ Myr for most disc mergers versus $\ga 1$ Gyr for most elliptical mergers), as a result of the higher central stellar density (attributed to increased star formation) and the more triaxial potential.
\item The exact SMBH merger time-scale in a galaxy merger is sensitive to the eccentricity when the SMBHs become hard.
\end{enumerate}
\end{description}

Given the results above, which demonstrate the sensitivity of the SMBH coalescing process to the galaxy properties, we can conclude that to improve the modelling of SMBH merger time-scales in galaxy mergers, it is critical to thoroughly consider AGN feedback and its subgrid implementations.

In this study, we have explored the impact of pure thermal and pure kinetic AGN feedback. In future work, we plan to take this a step further by considering a two-mode or more sophisticated AGN feedback model \citep[see discussions in e.g.][]{Naab2017}. In addition, it will be beneficial to expand our exploration to a broader parameter space by considering galaxy mergers with other properties of the initial galaxies (e.g. virial mass, SMBH mass, gas fraction, etc.) and various orbital configurations.

\section*{Acknowledgements}
We thank the anonymous referee for helpful comments. SL, DI, PHJ, FPR, and RJW acknowledge the support by the European Research Council via ERC Consolidator Grant KETJU (no. 818930). SL also acknowledges the supports by the National Natural Science Foundation of China (NSFC) grant (No. 11988101) and the K. C. Wong Education Foundation. PHJ and JMH acknowledge the support by the Academy of Finland grant no. 339127. TN acknowledges support from the Deutsche Forschungsgemeinschaft (DFG, German Research Foundation) under Germany’s Excellence Strategy -- EXC-2094 -- 390783311 from the DFG Cluster of Excellence ``ORIGINS''. AR acknowledges the support of the University of Helsinki Research Foundation. The numerical simulations used computational resources provided by the CSC -- IT Centre for Science, Finland. This research has made use of NASA’s Astrophysics Data System. We gratefully thank the developers of the open-source \textsc{python} packages that were used in the data analysis of this work, including \textsc{matplotlib} \citep{Hunter2007}, \textsc{numpy} \citep{Harris2020}, \textsc{scipy} \citep{Virtanen2020}, \textsc{astropy} \citep{astropy2013,astropy2018,astropy2022}, and \textsc{pygad} \citep{Rottgers2020}.

\section*{Data availability}
The simulation data used in this article will be shared upon a reasonable request to the corresponding author.

\bibliographystyle{mnras}
\bibliography{ref} 

\appendix

\section{Resolution study} \label{ap:res_study}

\begin{table*}
\begin{threeparttable}
\caption{Details of CoolStarKinAGN simulations with different resolutions. From left to right, the simulation name, the star particle number, the gas particle number, the dark matter particle number, the star (or gas) particle mass, the dark matter particle mass, the Plummer equivalent star softening length, the gas softening length, and the dark matter softening length.}
\label{tab:ap_sim_details}
\begin{tabular}{lccccccccccccc}
\hline
Simulations & $N_{\star}$ & $N_{\rm gas}$ & $N_{\rm DM}$ & $m_{\star}$ and $m_{\rm gas}$ & $m_{\rm DM}$ & $\epsilon_{\star}$ & $\epsilon_{\rm gas}$ & $\epsilon_{\rm DM}$ \\
 & [$10^{6}$] & [$10^{6}$] & [$10^{6}$] & [$10^{5}~{\rm M}_{\sun}$] & [$10^{5}~{\rm M}_{\sun}$] & [pc] & [pc] & [pc] \\
\hline
Low-res-$N_{\rm p}/4$ & 0.6 & 0.1 & 0.8 & 4.0 & 62.9 & 8.0 & 32.0 & 160.0 \\
Low-res-$N_{\rm p}/2$ & 1.2 & 0.2 & 1.6 & 2.0 & 31.0 & 6.0 & 25.0 & 126.0 \\
Fiducial-$N_{\rm p}$$^{\dagger}$ & 2.4 & 0.5 & 3.2 & 1.0 & 15.5 & 5.0 & 20.0 & 100.0 \\
Hi-res-$2N_{\rm p}$ & 4.9 & 0.9 & 6.4 & 0.5 & 7.7 & 4.0 & 16.0 & 80.0 \\
\hline
\end{tabular}
\begin{tablenotes}[flushleft]
 \footnotesize
 \item $^{\dagger}$ The fiducial simulations are the ones presented in the main text.
\end{tablenotes}
\end{threeparttable}
\end{table*}

\begin{figure*} 
\centering\includegraphics[width=400pt]{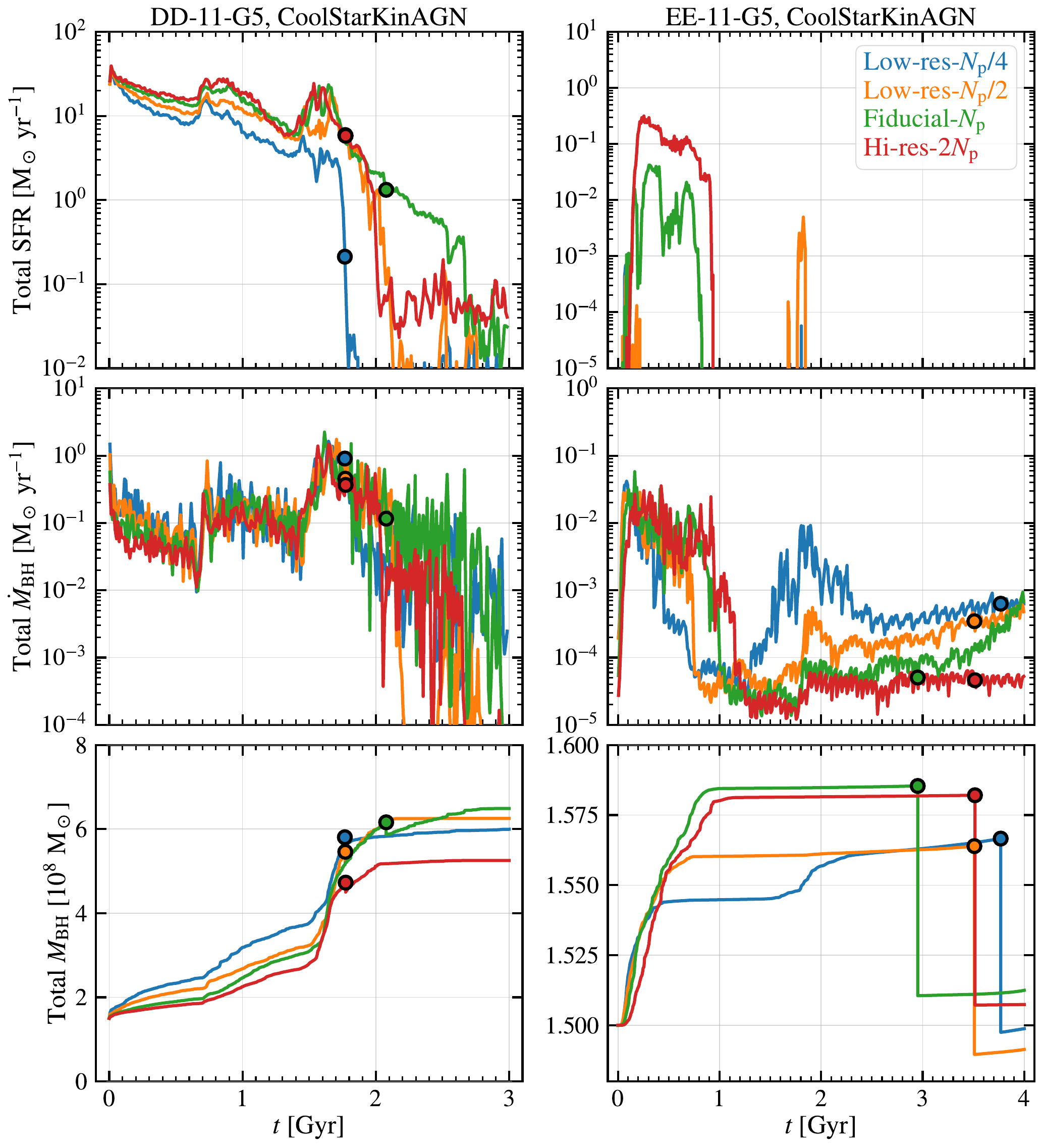}
\caption{Star formation and SMBH accretion histories from the CoolStarKinAGN runs with different resolutions. The DD-11-G5 and EE-11-G5 galaxy mergers are displayed in the left and right columns, respectively. From top to bottom rows, the total SFRs, the total SMBH accretion rates, and the total SMBH masses are plotted. The filled circles mark the SMBH merger events. Note that the SFRs and the SMBH accretion rates have been averaged over 10 Myr here. The simulations with different resolutions are distinguished using different colours. For both DD-11-G5 and EE-11-G5 mergers, the fiducial simulations (green), which are presented in the main text, show similar SFR and SMBH accretion histories as the high resolution runs (red), thus suggesting a relatively good resolution convergence.}
\label{fig:sfr_bhar_res_study}
\end{figure*}

\begin{figure*} 
\centering\includegraphics[width=400pt]{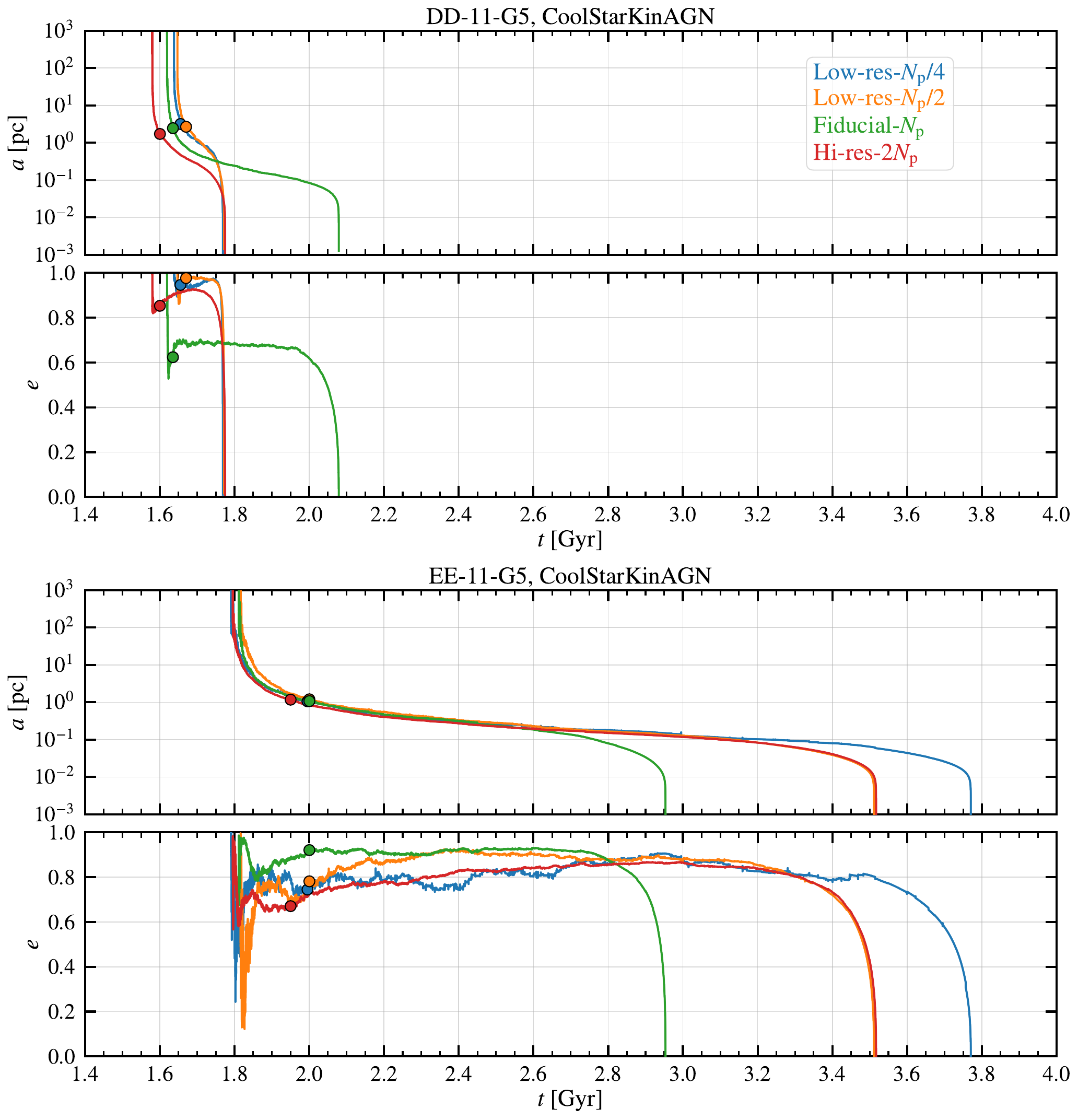}
\caption{Time-evolution of the SMBH binary orbital parameters from the CoolStarKinAGN runs with different resolutions. The DD-11-G5 and EE-11-G5 runs are shown in the upper and lower panels, respectively. From low to high resolutions, the runs are plotted with blue, orange, green, and red lines. The filled circles mark the time when the two SMBHs become hard in the different runs. Regardless of the numerical resolution, the disc galaxy mergers generally result in shorter SMBH merger time-scales compared to the elliptical galaxy mergers, with the exact time-scales being influenced by the eccentricities at $t_{\rm hard}$.}
\label{fig:orb_params_res_study}
\end{figure*}

To study the impact of the numerical resolution, we perform a resolution study with the kinetic AGN feedback model (i.e. CoolStarKinAGN), which was recently introduced into the {\sc ketju} code by \citetalias{Liao2023}, for both the DD-11-G5 and the EE-11-G5 galaxy mergers.

The runs with the fiducial resolution, which are presented and analysed in the main text, are named `Fiducial-$N_{\rm p}$' here. We further increase the resolution by a factor of 2 (`Hi-res-$2N_{\rm p}$') and decrease the resolution by a factor of 2 (`Low-res-$N_{\rm p}/2$') and 4 (`Low-res-$N_{\rm p}/4$'), to create simulations with different resolutions. The gravitational softening lengths are set according to $\epsilon = \epsilon_{\rm fid} / f^{1/3}$, where $\epsilon_{\rm fid}$ is the softening length of the fiducial run and $f \equiv N_{\rm p}/N_{\rm p, fid}$ is the particle number ratio between the resolution test run and the fiducial run. The detailed particle numbers, particle masses, and softening lengths of the different runs are summarized in Table~\ref{tab:ap_sim_details}. Only the softening lengths and accordingly the KETJU region radii are changed across the different resolution runs, all other simulation parameters remain identical to the fiducial ones.

The star formation and SMBH growth histories are plotted in Fig.~\ref{fig:sfr_bhar_res_study}. Overall, all the runs with different resolutions exhibit quite similar star formation and SMBH growth histories for both the DD-11-G5 and the EE-11-G5 mergers. The Low-res-$N_{\rm p}/4$ ones tend to have lower SFRs and higher SMBH accretion rates compared to the runs with higher resolutions. But the fiducial run shows results which are very similar to the Hi-res-$2N_{\rm p}$ run, suggesting a relatively good resolution convergence.

The time-evolution of the orbital parameters are summarized in Fig.~\ref{fig:orb_params_res_study}. As discussed in the main text, the exact SMBH merger time-scale of an individual simulation is strongly influenced by $e_{\rm hard}$. However, regardless of the numerical resolution, the disc galaxy mergers generally result in shorter SMBH merger time-scales compared to elliptical galaxy mergers, which is attributed to the higher central stellar densities in the disc galaxy merger remnants.

To conclude, our runs with the kinetic AGN feedback exhibit a relatively good convergence across different resolution tests.

\section{Influence of AGN feedback on isolated elliptical galaxies}\label{ap:isolate_ETG}

\begin{figure} 
\centering\includegraphics[width=\columnwidth]{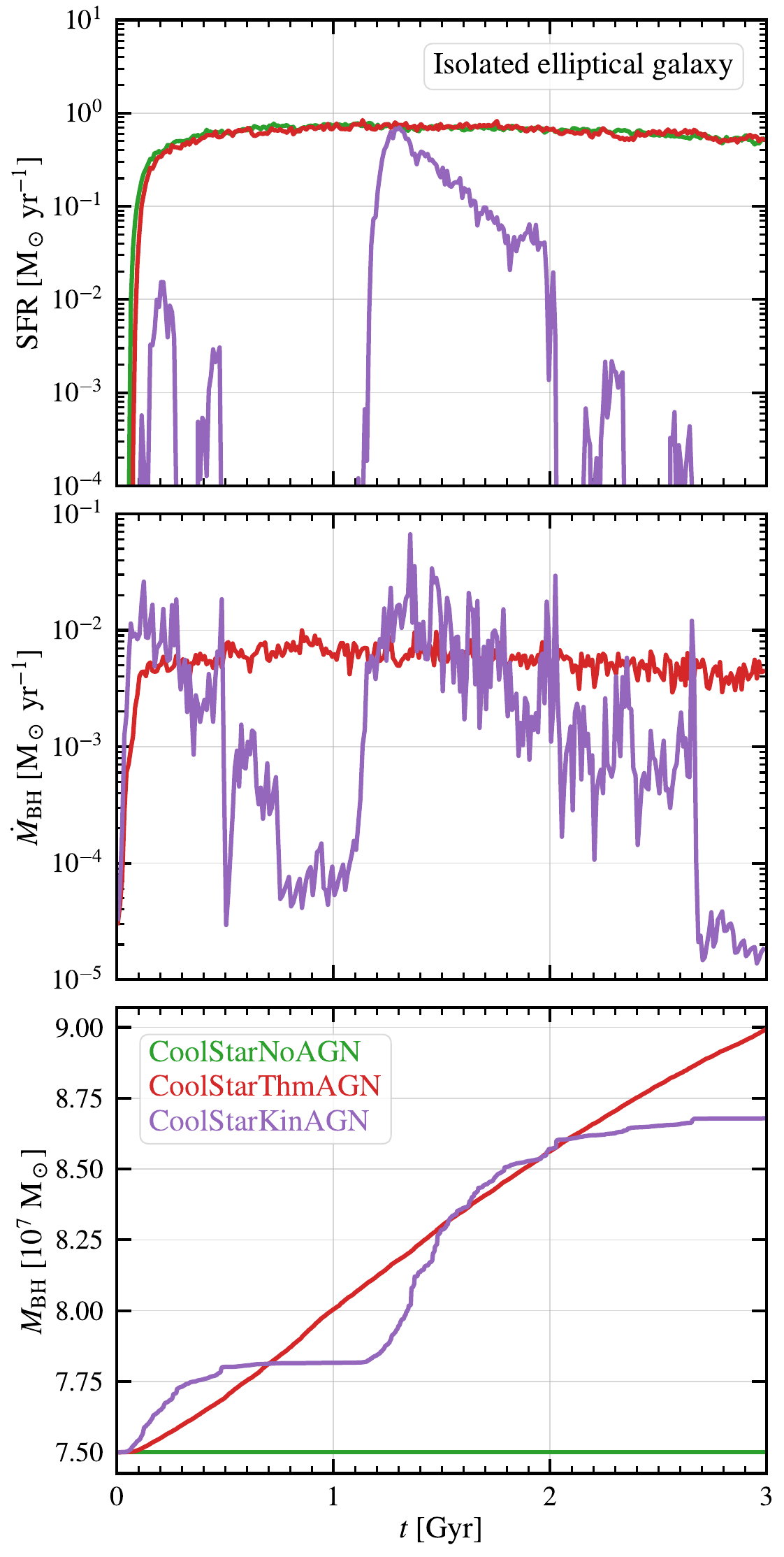}
\caption{Star formation and SMBH accretion histories in the isolated elliptical galaxy simulation. From top to bottom panels, the SFR, SMBH accretion rate, and SMBH mass as a function of simulation time are plotted. The CoolStarNoAGN, CoolStarThmAGN, and CoolStarKinAGN runs are shown with green, red, and purple, respectively.}
\label{fig:isolate_sfr_bhar}
\end{figure}

\begin{figure} 
\centering\includegraphics[width=\columnwidth]{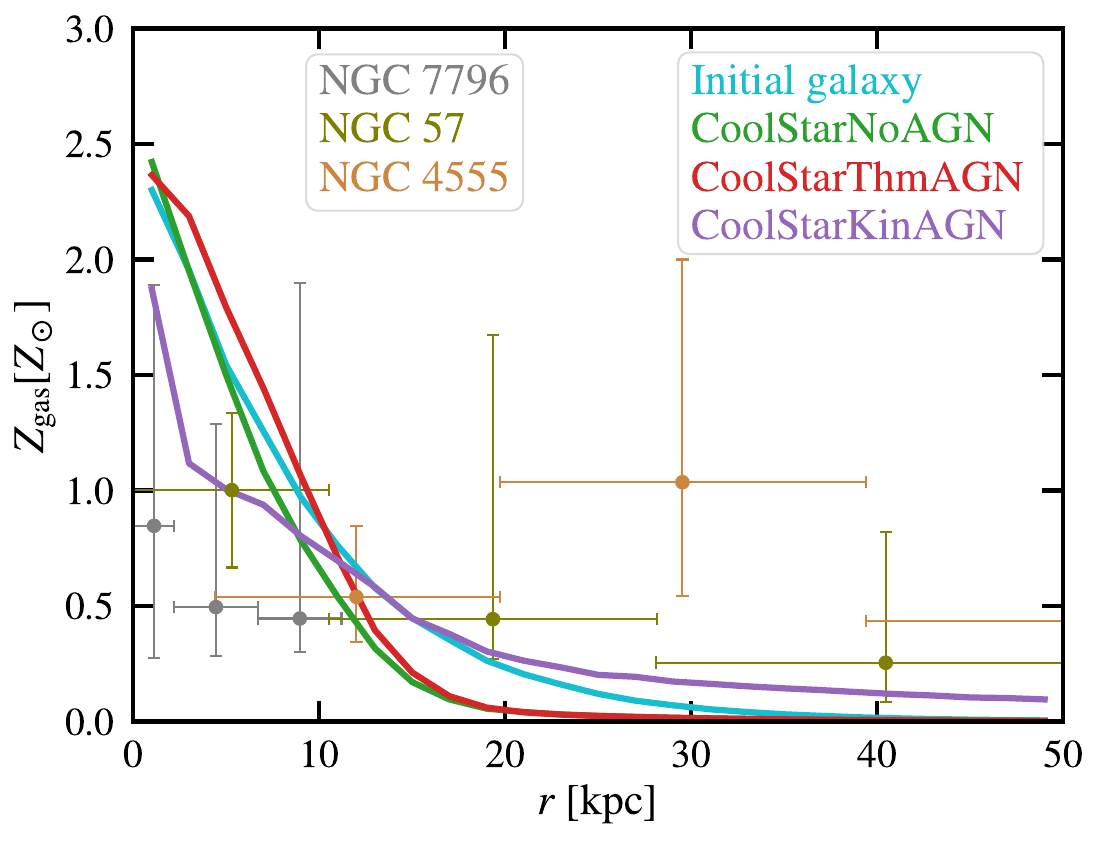}
\caption{Radial profiles of the gas metallicity from the isolated elliptical galaxy runs. The profiles for the initial galaxy and the evolved galaxies at $t = 3$ Gyr from the CoolStarNoAGN, CoolStarThmAGN, and CoolStarKinAGN runs are plotted with cyan, green, red, and purple, respectively. The observed gas metallicity profiles for isolated (i.e. residing in low-density environment) elliptical galaxies NGC 7796 and NGC 57 \citep{OSullivan2007} and that for NGC 4555 \citep{OSullivan2004} are overplotted for comparison (see also \citealt{Eisenreich2017}).}
\label{fig:gas_metal_profile}
\end{figure}

It was showed in \citet{Eisenreich2017} that the details of the AGN feedback implementations have a significant influence on the evolution and properties of isolated elliptical galaxies. In this appendix, we perform similar simulations to study how our AGN feedback models affect the quiescence and the circumgalactic medium metal enrichment of isolated elliptical galaxies.

We adopt the E1 galaxy as the initial condition and evolve it for 3 Gyr using the CoolStarNoAGN, CoolStarThmAGN, and CoolStarKinAGN simulation setups. The simulation parameters used in these isolated galaxy runs are identical to those of the galaxy merger runs as detailed in Section~\ref{subsec:sim}.

The star formation and SMBH growth histories from different runs are plotted in Fig.~\ref{fig:isolate_sfr_bhar}. Without AGN feedback, the SFR settles to a value of ${\sim} 1~{\rm M}_{\sun}{\rm yr}^{-1}$, or equivalently the specific SFR (sSFR $\equiv {\rm SFR} / M_{\star}$) is maintained at ${\sim} 10^{-11}{\rm yr}^{-1}$, making the galaxy a green valley galaxy, which lies between the blue star-forming and red quiescent sequences \citep{Salim2014}. Star formation occurs within the circumnuclear disc, which forms as gas cools and condenses from the rotating hot gas halo (i.e. cooling flows). In the CoolStarThmAGN run, the energy from the thermal AGN feedback heats the gas close to the SMBH, but the thermal energy quickly radiates away due to the efficient cooling in the circumnuclear disc. Consequently, thermal AGN feedback only affects the star formation in a small central region, and the total SFR as a function of time is quite close to that of the CoolStarNoAGN run, resulting in a green valley galaxy. However, in the CoolStarKinAGN run, we observe that for the majority of the simulation period, the total SFR remains below $0.1~{\rm M}_{\sun}{\rm yr}^{-1}$, which results in the galaxy being categorized as a red and quiescent galaxy. The strong pulsed kinetic AGN feedback effectively expels gas particles from the galaxy centre, causing collisions with gas particles at larger radii and thus heating the surrounding gas. This mechanism efficiently removes gas from the centre, preventing cooling flows and promoting a quiescent galaxy state. The SMBH accretion history demonstrates clear self-regulated behaviour, wherein an increased gas availability for SMBH accretion triggers stronger kinetic AGN feedback, rapidly removing the excess gas and reducing the accretion rate. Consequently, the SMBH mass exhibits step-like growth patterns.

The gas metallicity profiles of the evolved galaxies at $t = 3$ Gyr from the different runs are displayed in Fig.~\ref{fig:gas_metal_profile}. Compared to the initial metallicity profile, the CoolStarNoAGN profile tends to be more concentrated as a result of gas cooling and the enrichment from the very central new stars. Once the thermal AGN feedback is included, the profile becomes slightly more extended than the CoolStarNoAGN one. On the other hand, the kinetic AGN feedback is much more effective in lowering the central gas metallicity and flattening the profile, which agrees better with the profiles of the observed isolated elliptical galaxies plotted in Fig.~\ref{fig:gas_metal_profile}.

To summarize, compared to thermal AGN feedback, kinetic AGN feedback is more effective in driving gas out of the galaxy centre, quenching star formation, and flattening the gas metallicity profile. Overall, the results of our isolated elliptical galaxy tests are in line with the findings in \citet{Eisenreich2017}.

\bsp
\label{lastpage}

\end{document}